\documentclass[aps,prx,reprint,superscriptaddress,showpacs]{revtex4-1}
\usepackage{graphicx}
\UseRawInputEncoding 
\usepackage{amssymb}
\usepackage{color}
\usepackage{amsmath}
\usepackage{ulem}
\usepackage{mathrsfs}

\newcommand{\up}{\uparrow}
\newcommand{\dn}{\downarrow}

\newcommand{\be}{\begin{eqnarray}}
\newcommand{\ee}{\end{eqnarray}}
\newcommand{\la}{\langle}
\newcommand{\ra}{\rangle}

\begin{document}

\title{Improvements to the Stochastic Series Expansion method for the $JQ_2$ model with a magnetic field}

\author{Lu Liu}
\affiliation{School of Physics, Beijing Institute of Technology, Beijing 100081, China}
\email{liulu96@bit.edu.cn}

\begin{abstract}
   The Stochastic Series Expansion (SSE) quantum Monte Carlo method with directed loops is very efficient for spin and boson systems. The Heisenberg model and its generalizations, such as the $JQ_2$ model, are extensively simulated via this method. When introducing magnetic field in these models, the SSE method always combines the field with the diagonal part of the Heisenberg interactions (${S_i}^z{S_j}^z$) and take them as the new diagonal operators. In general, this treatment is reasonable. However, when studying Hamiltonians which have other interactions or even don't contain the Heisenberg interactions, this general treatment will not be efficient or even not work. We suggest that when doing directed-loop simulations, the magnetic field can be put into other interactions. This new treatment, in some cases, improves the simulation efficiency. Using the $JQ_2$ model with magnetic field as an example, we here demonstrate this new SSE method. Such new treatment significantly improves the efficiency when the $Q_2$ interactions are large. The autocorrelations are reduced a lot compared to the previous approach. In addition, we argue that we can divide the magnetic field into two parts and combine them with both the $J$ and $Q$ operators respectively. This treatment also improves the simulation efficiency. 
The underlying mechanism is that these two new SSE methods can utilize the main part or even all part of operators in operator products to do the directed-loop updates. Such idea can also be applied to other models with magnetic field and it will speed up the simulations. 

\end{abstract}

\date{\today}

\maketitle

\section{Introduction}
\label{sec:intro}
    The Stochastic Series Expansion (SSE) quantum Monte Carlo (QMC) method \cite{sse1} with loop update \cite{sse2,sse-loop,sse-direct} is a very powerful method for quantum spin systems and boson systems. The Heisenberg model and its generalizations (e.g. the $JQ$ models) are extensively simulated by this method. When introducing magnetic field, Sylju\r{a}sen and Sandvik proposed SSE method with directed-loop update which includes the field operators into the Heisenberg interactions \cite{sse-direct}. The Heisenberg model is taken as a specific example to demonstrate that the directed-loop simulations are very efficient for the full range of magnetic field (zero to the saturation point). When simulating the generalizations of the Heisenberg model which not only contain the Heisenberg interactions but also have other interactions, people in general still combine the magnetic field with the Heisenberg interactions \cite{iaizzi2017,iaizzi2018,cuiyi}. Such treatment does work for most cases. However, it is not the most efficient way for some extreme cases. In this paper, we argue that not only can the Heisenberg interaction be combined with the external magnetic field, but also other interactions can contain the magnetic field. The simulations, in some cases, are more efficient when we put the magnetic operators into other interactions. What's more, we can divide the magnetic field and combine them with all the types of interactions. 
In this paper, we consider the $JQ_2$ model with magnetic field as a specific example. We present that the new versions of the SSE method with directed loops are more efficient than that of the general SSE method when $Q$ is large.

This paper proceeds as follows. In Sec. \ref{sec:jq2model}, backgrounds on the $JQ_2$ model are given. In Sec. \ref{sec:sseconf}, we introduce the basic concepts of the SSE method. In Sec. \ref{sec:general}, we briefly review the general SSE method with directed loops for the $JQ_2$ models with an external magnetic field. In Sec. \ref{sec:modify}, we present our two new versions of the SSE methods for the $JQ_2$ model with magnetic field. In Sec. \ref{sec:simu}, we present simulation results and show that the new versions of the SSE method do work and decrease the autocorrelation times significantly. We conclude in Sec. \ref{sec:summary}.

\section{The $JQ_2$ model}
\label{sec:jq2model}
  The $JQ_2$ model is a generalization of the Heisenberg model, which has four-spin interactions $Q$ on every plaquette.
  The Hamiltonian is expressed as
   \be
   H&=&-J\sum_{\la ij\ra}P_{ij}-Q\sum_{\la ij,kl\ra}P_{ij}P_{kl}\\
    &=&-(H_J+H_Q)\nonumber
   \ee
 where $P_{ij}=1/4-{\bf S_i}\cdot{\bf S_j}$ is the singlet projector operator, $\la ij\ra$ represent two nearest neighbor sites, $\la ij,kl\ra$ represents four sites on one plaquette and the index pairs $ij$, $kl$ form two parallel bonds on horizontal or vertical directions. The summations are over all nearest neighbors for the $J$ terms and all translations of the vertical and horizontal stacks for the $Q$ terms. $H_J$ and $H_Q$ stand for the $J$ and $Q$ terms in the Hamiltonian. In this paper, we take $J=1$ as the unit of energy. We call every interaction as a bond and the $JQ_2$ model has two types of bonds: the $J$ bonds and $Q$ bonds. This model was proposed by Sandvik to show the deconfined quantum phase transition from the N\'eel state to the valence-bond-solid (VBS) state \cite{jq2} and later other variants of the $JQ$ models were proposed, such as the $JQ_n$ model (the $JQ_n$ model means every $Q$ operator is the products of $n$ singlet projector operators), the checkerboard $JQ$ ($CBJQ$) model.
Lots of novel properties were found in these models. In the $JQ_3$ model, there also exists the deconfined quantum phase transition \cite{jq3}; there is a symmetry enhanced first-order phase transition in the $CBJQ$ model \cite{cbjq-bowen}; in the $JQ_6$ model, an emergent $SO(5)$ symmetry was observed \cite{jq6}; in a modulated$-J$ $JQ$ model, the multicritical deconfined quantum criticality was found \cite{jq-bowen}.
 
There is a deconfined quantum phase transition in the $JQ_2$ model. When $g=J/Q$ is large, the ground state is the N\'eel state and when $g=J/Q$ is small enough, the ground state is the VBS state. The transition point between these two phases is $g_c=J/Q\approx0.045$, which means when we set $J=1$, the value of $Q_c$ will be around $22$ \cite{huishao-science}. Such value is really large. When we are interested in the properties of the $JQ_2$ model around $Q_c$ or the properties of VBS state, we have to set $Q$ very large. The SSE method is efficient for the simulations of this model. However, when introducing an external magnetic field, the situation will be different. We find the general SSE method with directed loops, where the magnetic field is combined with the Heisenberg interactions ($J$ terms), is less efficient. The autocorrelation time is much longer and we need very large Monte Carlo steps to obtain high-quality data. When the system size and inverse temperature increase, the autocorrelation time will increase significantly and we can even not get the correct results because of the limitation of computational resources. An intuitive idea is that for the $JQ_2$ model without external field, the average number of times that the operators of the $J$ ($Q$) bonds appearing in the operator string is proportional to the expectation value of $H_J$ ($H_Q$). When $Q$ is much larger than $J$, the expectation value of $H_Q$ will also be much larger than that of $H_J$. It means the $Q$ bonds occur more frequently than the $J$ bonds in the operator strings (see Appendix. \ref{app:number}). When introduced magnetic field, the general directed-loop update only makes use of the small part ($J$ bonds) of the operator strings, which of course is not efficient. In this article, we propose that we can use the main part of the operator string ($Q$ bonds) to do the directed-loop updates or we can use all the bonds in the operator strings (both $J$ and $Q$ bonds) to do the directed-loop updates. The two treatments reduce the autocorrelation time significantly and make the simulations more efficient.

In the next two sections, we will introduce the general concepts of SSE method and talk about the general treatment of the $JQ_2$ model with magnetic field in the SSE method.

\section{Basic concepts of SSE method}
\label{sec:sseconf}
In this section, we will introduce some basic concepts of the SSE method. 

For general models with $N$ spins, such as the Heisenberg antiferromagnet and the $JQ$ model, we use the standard basis for these models
\be
|\alpha\ra=|S_1^z,S_2^z,\cdots,S_N^z\ra
\ee
and write the Hamiltonians in terms of bond operators $H_b$,
\be
H=-\sum_{b=1}^{N_b}H_b
\ee
 where every index $b$ refers to one interaction term: for the Heisenberg interactions, one bond only contains two spins and for multispin interactions, such as $Q$ terms in the $JQ$ model, one $Q$ bond can contain more spins. $N_b$ is the number of bonds. In order to carry out the SSE simulations, we need divide every bond into $\mathbb{N}$ operators: 
\be
H_b=H_{1,b}+\sum_{i=2}^{\mathbb{N}}H_{i,b}
\label{hbond}
\ee
where $H_{1,b}$ is the diagonal part of bond operator and $H_{i,b} (i\geq 2)$ are the off-diagonal bond operators. We will see that for the Heisenberg interaction $\mathbb{N}=2$ and for the $Q_n$ interaction $\mathbb{N}=2^n$ (the $Q_2$ interactions $\mathbb{N}=4$; the $Q_3$ interactions, $\mathbb{N}=8$). The definition of the $JQ$ model has been shown in Sec. \ref{sec:jq2model}.

The starting point of the SSE method is the Taylor expansion of the partition function:
\be
Z={\rm Tr} \{{\rm e}^{-\beta H}\}=\sum_\alpha\sum_{n=0}^\infty \frac{(-\beta)^n}{n!}\la\alpha|H^n|\alpha\ra
\ee
where the trace is written as summation of basis ${|\alpha\ra}$ and $\beta=1/T$ is the inverse temperature. Next according to Eq. (\ref{hbond}), the operator string $H^n$ need to be expanded as summations of products of diagonal and off-diagonal bond operators. We truncate the expansion power $n$ at a maximum value $M$ and then fix the operator products length as $M$ by introducing some unit operators. For general expansion power $n$ $(n\leq M)$, we need insert $M-n$ unit operators $H_{0,0}=I$ in the operator products in all possible ways. Finally, the partition function is written as
\be
Z=\sum_\alpha\sum_{S_M}\frac{\beta^n(M-n)!}{M!}\la \alpha|\prod_{i=1}^M H_{a_i,b_i}|\alpha\ra
\label{part}
\ee
where $n$ is the number of non-unit bond operators, $a_i=1,2,\cdots,\mathbb{N}$ corresponds the type of operators ($0$, unit; $1$, diagonal; $2,3,\cdots$, off-diagonal) and $b_i=0,1,2,\cdots,N_b$ is the bond index ($0$ for unit operators, $1,2,\cdots,N_b$ for non-unit bonds). $S_M$ is the configurations of operator products. Such a product can be referred to by an operator-index sequence
\be
S_M=[a_1,b_1],[a_2,b_2],\cdots,[a_M,b_M].
\ee
For simplicity, we sometimes use the notation $[a,b]_p$ to represent $[a_p,b_p]$, where $p$ can be thought as the index of imaginary time.

One can show that the average expansion order is
\be
\la n\ra=\beta|E|
\ee
where $E$ is the system energy, $E=\la H\ra$ \cite{sse-loop,sse-direct}. The width of the expansion order is approximately $\la n\ra^{1/2}$. The cutoff $M$ can be chosen so that $n$ can never reaches this value. The truncation error is then negligible.

The Monte Carlo simulation can be started with some random state $|\alpha\ra$ and ``unit" operator string $S_M=[0,0]_1,[0,0]_2,\cdots,[0,0]_M$. The general SSE sampling of configurations $(\alpha, S_M)$ contains two different types of updates, which ensure the ergodicity of the sampling. The first update (diagonal update) is of the update between the unit operator $[0,0]_p$ and the diagonal operator $[1,b]_p$. Such update will change the expansion order $n$ by $\pm 1$. The corresponding Metropolis acceptance probabilities are
 \be
 P([0,0]_p\rightarrow[1,b]_p)={\rm min}\left\{1,\frac{N_b\beta\la\alpha(p)|H_{1,b}|\alpha(p)\ra}{M-n}\right\}\\
 P([1,b]_p\rightarrow[0,0]_p)={\rm min}\left\{1,\frac{M-n+1}{N_b\beta\la\alpha(p)|H_{1,b}|\alpha(p)\ra}\right\}
 \ee
The second update (off-diagonal update) is of the update between the diagonal operators $[1,b]_p$ and the off-diagonal operators $[a,b]_p$, where $a\geq2$. It can be done by the loop update. People can refer to article \cite{sse-loop,sse-direct} for detail information about the off-diagonal update for the Heisenberg model. These two new SSE methods introduced in this paper have the same update processes, which also contain both the diagonal update and the off-diagonal update.

It is convenient to define a Monte Carlo step (MCS) for the SSE simulation. One MCS contains a sweep of diagonal updates at all imaginary time positions and then do the construction of linked list. After this construction, a fixed number of loop updates are applied. Thus every MCS contains both types of updates.

During the simulation, we should firstly evolve the initial configuration to the equilibrated configurations and then the reliable measurements are possible. Thus in Monte Carlo simulations, one firstly do some ``equilibration" MCSs and then do some ``measure" MCSs. The number of these two MCSs depends on the equilibrium correlation time and the autocorrelation time, we will not discuss this in detail. The physical observables are measured during the ``measure" MCSs. A general observable $A$ (mostly diagonal observables) can be measured according to
\be
\la A\ra&=&\frac{1}{Z}\sum_{\alpha,S_M}\frac{\beta^n(M-n)!}{M!}\la \alpha|A\prod_{i=1}^M H_{a_i,b_i}|\alpha\ra\nonumber\\
&=&\sum_{\alpha,S_M}A(\alpha,S_M)W(\alpha,S_M)/\sum_{\alpha,S_M}W(\alpha,S_M)\nonumber
\ee
where 
\be
A(\alpha,S_M)&=&\frac{\la \alpha|A\prod_{i=1}^M H_{a_i,b_i}|\alpha\ra}{\la \alpha|\prod_{i=1}^M H_{a_i,b_i}|\alpha\ra}\nonumber\\
W(\alpha,S_M)&=&\frac{\beta^n(M-n)!}{M!}\la \alpha|A\prod_{i=1}^M H_{a_i,b_i}|\alpha\ra\nonumber
\ee

We will not discuss the measurement in detail here. Several observables have been derived in Ref. \cite{sse-loop,obs}. The off-diagonal correlation functions have been studied in Ref. \cite{obs-off}. 

\section{General SSE method for the $JQ_2$ model with magnetic field}
\label{sec:general}
In this section, we introduce the general SSE method that deals with the magnetic field in the $JQ_2$ model. 
The Hamiltonian of the $JQ_2$ model can be written as
\be
H_{JQ_2-h}=-\sum_{\la ij\ra}JP_{ij}-\sum_{\la ij,kl\ra}QP_{ij}P_{kl}-h\sum_i S_i^z
\label{jq2h1}
\ee
The last term in Eq. (\ref{jq2h1}) is the magnetic field. When doing the SSE simulations on this model, we generally put the magnetic field into the diagonal part of the $J$ terms. In order to ensure all matrix elements of the new diagonal part of the $J$ terms are not negative, we also need to add a constant. Such constant is not unique, people can choose any value as long as the new diagonal part of the $J$ terms have no negative matrix elements. 
The final Hamiltonian of the $JQ_2$ model with magnetic field is written as
\be
H_{JQ_2-h}=&-&\sum_{\la ij\ra}J(P_{ij}+h_b(S_i^z+S_j^z+1)+\epsilon)\nonumber\\
&-&\sum_{\la ij,kl\ra}QP_{ij}P_{kl}\label{eq:jq2h}
\ee
In Eq. (\ref{eq:jq2h}), we have defined the magnetic field on a $J$ bond, the field strength is $h_b=h/zJ$ and $z$ is the coordination number. In addition, we have added a constant $h_b+\epsilon$ on every $J$ bond and $\epsilon\geq 0$. 

\begin{figure}[t]
  \includegraphics[width=75mm,clip]{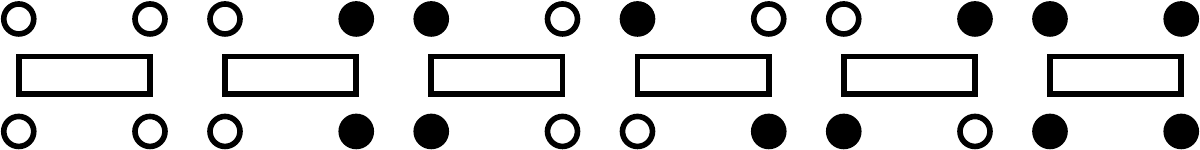}
  \caption{$6$ different vertices for the $J$ operators. The horizontal bar represents the operators. The circles beneath (above) represent the spin state (open and solid circles for spin-$\up$ and spin-$\dn$, respectively) before (after) operation with the $J$ operators. We denote these six vertices as $\Gamma_i, i=1,2,3,4,5,6$. We denote the four circles as four legs of the vertex.}
  \label{confj}
  \end{figure}

When doing the SSE simulations, we divided every $J$ term in Eq. (\ref{eq:jq2h}) into two operators: one is diagonal $H^{(1)}_{ij}=1/4-S_i^zS_j^z+h_b(S_i^z+S_j^z+1)+\epsilon$ and the other is off-diagonal $H^{(2)}_{ij}=-(S_i^xS_j^x+S_i^yS_j^y)=-1/2(S_i^+S_j^-+S_i^-S_j^+)$. 
Every matrix element of these two types of operators can be represented by a vertex and they will give six different vertices. We denote these vertices as $\Gamma_i$ $(i=1,2,3,4,5,6)$. The six different vertices contain four diagonal vertices and two off-diagonal vertices (In diagonal vertices, the spin state doesn't change after the operation of a operator. In the off-diagonal vertices, the spin state will change after the operation of a operator). The four diagonal vertices are $\Gamma_1:\la\up\up|H^{(1)}|\up\up\ra$, $\Gamma_2:\la\up\dn|H^{(1)}|\up\dn\ra$, $\Gamma_3:\la\dn\up|H^{(1)}|\dn\up\ra$, $\Gamma_6:\la\dn\dn|H^{(1)}|\dn\dn\ra$ and the two off-diagonal vertices are $\Gamma_4:\la\up\dn|H^{(2)}|\dn\up\ra$, $\Gamma_5:\la\dn\up|H^{(2)}|\up\dn\ra$. The index of these six vertices can be set arbitrarily. The weight of these six vertices (matrix element of operators) are $W(\Gamma_1)=2h_b+\epsilon$, $W(\Gamma_2)=1/2+h_b+\epsilon$, $W(\Gamma_3)=1/2+h_b+\epsilon$, $W(\Gamma_4)=1/2$, $W(\Gamma_5)=1/2$, $W(\Gamma_6)=\epsilon$ respectively. These vertices are shown in Fig. (\ref{confj}).

Every $Q_2$ term contains a product of two singlet projector operators. In the SSE simulations, it also should be divided.
As every singlet projector operator $P_{ij}$ can be divided into two part: one is diagonal $H^{(1)}_{ij}=1/4-S_i^zS_j^z$ and the other is off-diagonal $H^{(2)}_{ij}==-1/2(S_i^+S_j^-+S_i^-S_j^+)$. Note that the diagonal operator $H^{(1)}$ here, which doesn't contain magnetic field part, is different from the diagonal operator of $J$ terms mentioned above. However we denote both of them as $H^{(1)}$ for simplicity. Thus for every $Q_2$ term, it can be divided into four different operators: $H_1=H^{(1)}_{ij}H^{(1)}_{kl}$, $H_2=H^{(1)}_{ij}H^{(2)}_{kl}$, $H_3=H^{(2)}_{ij}H^{(1)}_{kl}$ and $H_4=H^{(2)}_{ij}H^{(2)}_{kl}$. $H_1$ is the diagonal part of the $Q$ operator and the three others are off-diagonal parts of the $Q$ operator. Every off-diagonal term of the $Q_2$ operators has at least one $H^{(2)}$. We denote these four terms as
$H_1=Q^{(11)}_{ijkl}, H_2=Q^{(12)}_{ijkl}, H_3=Q^{(21)}_{ijkl}, H_4=Q^{(22)}_{ijkl}$.



For $JQ_2$ model with magnetic field, every diagonal part of the $J$ operators contains the field and every diagonal part of the $Q$ operators doesn't contain any magnetic field. Thus when doing the diagonal update, the weight for the $J$ diagonal operators should include the field operators. The off-diagonal update is a bit complicated. When a loop encounters the $J$ operators, the loop should choose the exit leg with a probability which corresponds to the particular solution of the directed-loop equations (see Eq. ($29$),($31$) in Ref. \cite{sse-direct}). When the loop encounters the $Q$ operators, it will just do the simple switch-and-reverse moves. This treatment introduced above is the general SSE method. In the next section, we will present a new version of SSE method for the $JQ_2$ model with magnetic field.

\section{Modified version of SSE method for the $JQ_2$ model with magnetic field}
\label{sec:modify}
As mentioned above, the critical point $Q_c$ is really large for the $JQ_2$ model. When $Q$ is large, the $J$ operators appear much less frequently than the $Q$ operators. If we study the properties of the $JQ_2$ model at large $Q$ with external magnetic field, we will find the efficiency of the general SSE method with directed loops is very poor and we suggest a modified version of the SSE method, which is more efficient.

As the operators appearing in the operator products will mostly be the $Q$ operators, if we can do the SSE simulations by combining the $Q$ operators and magnetic field together, the simulations is more efficient than the general SSE method. In this section we will show such combination can indeed be applied in the SSE method.  What's more, if we only combine the magnetic field with the $Q$ operators, the directed-loop updates only make use of the $Q$ operators. The $J$ operators don't participate in the directed-loop updates. When a loop encounters the $J$ operators, it just does the switch-and-reverse moves.
We argue that the magnetic field can be divided into two parts. The first part is combined with the $J$ operators (just as the general method) and the second part is combined with the $Q$ operators. Any proportion of the division works for the SSE method, but the best proportion depend on the parameters (the strength of the $J$ and the $Q$ interactions).
 As all non-unit operators will take participate in the directed-loop updates, it speed up the simulations much more if the proportion of division is chosen properly. 
In order to distinguish these three different SSE methods, we denote the general SSE method as ``J-SSE" (the directed loops only work on the $J$ operators). The modified SSE method, in which the magnetic field is combined with the $Q$ operators, is denoted as ``Q-SSE". The final modified SSE method, in which the magnetic field is split and combined to both $J$ and $Q$ operators, is denoted as ``JQ-SSE". The ``J-SSE" and ``Q-SSE" method can be thought as two extreme cases of ``JQ-SSE" method.

In this section, we present how to combine the magnetic field with the $Q$ operators in the modified ``Q-SSE" and ``JQ-SSE" methods. We then show how to do the simulations in the ``Q-SSE" method. The correctness of  the ``Q-SSE" method will be proven in the next section by comparing the results of the ``Q-SSE" method with the exact diagonalization (ED) method. 

The Hamiltonian of the $JQ_2$ model with magnetic field can also be written as
\be
H^\prime_{JQ_2-h}=&-&\sum_{\la ij\ra}JP_{ij}\nonumber\\
&-&\sum_{\la ij,kl\ra}Q(P_{ij}P_{kl}+h_q(S_i^z+S_j^z+S_k^z+S_l^z))\nonumber
\ee
where have defined the magnetic field on a $Q$ bond and the strength is $h_q=h/2zQ$. 

We then put the magnetic field term $h_q(S_i^z+S_j^z+S_k^z+S_l^z)$ into the diagonal part of the $Q_2$ operator: $Q^{(11)}_{ijkl}$. The diagonal part now is written as
\be
Q^{(11)}_{ijkl}=H^{(1)}_{ij}H^{(1)}_{kl}+h_q(S_i^z+S_j^z+S_k^z+S_l^z)\nonumber
\ee

In order to make all matrix elements (vertices weight) not negative for the diagonal part of the $Q$ operators, we also need to add a constant as that in the traditional SSE method (``J-SSE") for the diagonal part of the $J$ operators. The constant we choose in this paper is $2h_q$ for every $Q$ bond and of course this constant is also not unique. People can choose another constant and the principle of derivation is the same. The constant $2h_q$ chosen here corresponds to $\epsilon=0$ for the ``J-SSE" method. In this paper, we choose $\epsilon=0$ for the ``J-SSE" method and ``JQ-SSE" method in order to focus on the efficiency difference when combining the magnetic fields with different types of operators. Finally the diagonal part of the $Q_2$ operators is written as
\be
Q^{(11)}_{ijkl}=H^{(1)}_{ij}H^{(1)}_{kl}+h_q(2+S_i^z+S_j^z+S_k^z+S_l^z)\nonumber
\ee

In our simulations, the finial Hamiltonian now is written as
\be
H_{JQ_2-h}=&-&\sum_{\la ij\ra}J(H_{ij}^{(1)}+H_{ij}^{(2)})\nonumber\\
&-&\sum_{\la ij,kl\ra}Q(Q^{(11)}_{ijkl}+Q^{(12)}_{ijkl}+Q^{(21)}_{ijkl}+Q^{(22)}_{ijkl})\nonumber\\
\ee
where $H_{ij}^{(1)}=1/4-S_i^zS_j^z, H_{ij}^{(2)}=-1/2(S_i^+S_j^-+S_i^-S_j^+)$ and $Q^{(11)}_{ijkl}=H^{(1)}_{ij}H^{(1)}_{kl}+h_q(2+S_i^z+S_j^z+S_k^z+S_l^z)$, $Q^{(12)}_{ijkl}=H^{(1)}_{ij}H^{(2)}_{kl}$, $Q^{(21)}_{ijkl}=H^{(2)}_{ij}H^{(1)}_{kl}$ and $Q^{(22)}_{ijkl}=H^{(2)}_{ij}H^{(2)}_{kl}$.

\begin{figure*}[t]
   \includegraphics[width=160mm,clip]{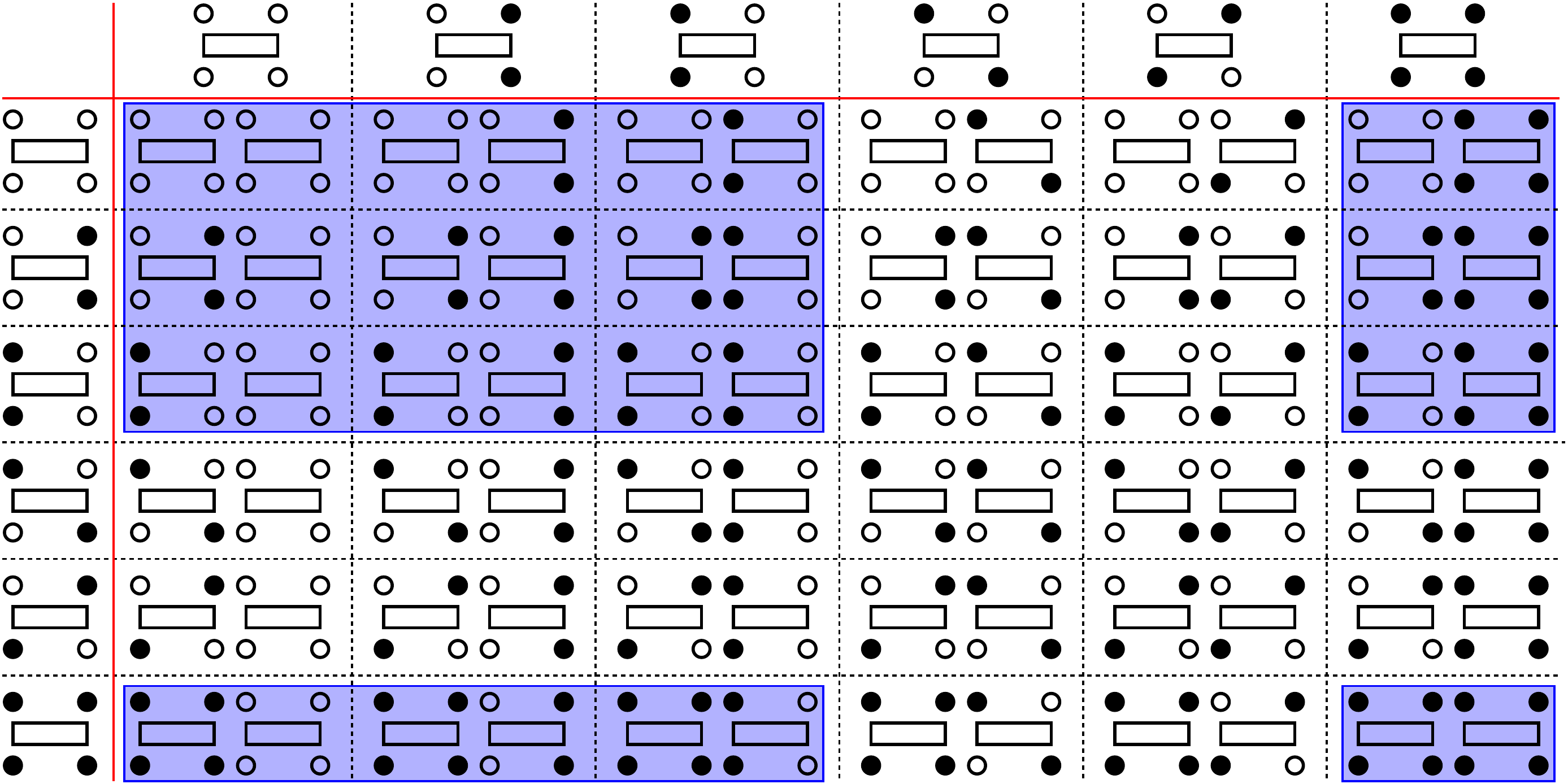}
   \caption{$36=6\cdot 6$ types of vertices for the $Q_2$ operators. The diagonal vertices are shown in blue regions and the others are off-diagonal vertices.}
   \label{confq2}
   \end{figure*}

As every $Q$ operator can be thought as the product of two $J$ bonds, thus every $Q$ operator will give $36=6\cdot6$ different types of vertices: $16$ of them are diagonal, $20$ of them are off-diagonal. Every vertex of the $Q$ bond acts on $4$ sites and has $8$ legs. The $36$ kinds of vertices are shown in Fig. (\ref{confq2}). The $16$ kinds of diagonal vertices are shown in blue regions and the $20$ kinds of off-diagonal vertices are shown in white region.

    \begin{figure*}[t]
    \includegraphics[width=160mm,clip]{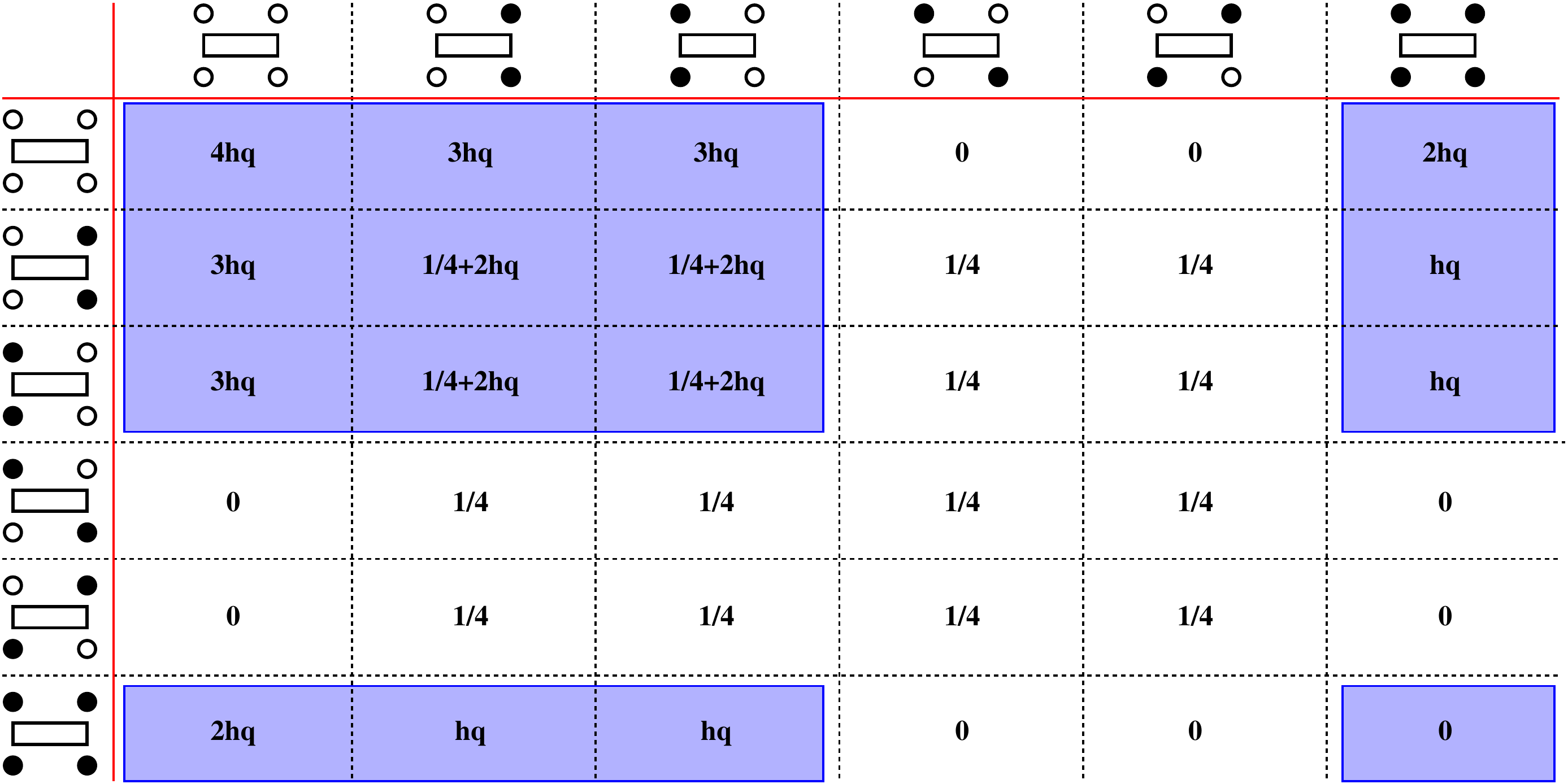}
    \caption{The weight of $36=6\cdot 6$ types of vertices for the $Q_2$ operators. The diagonal vertices are shown in blue regions and the others are off-diagonal vertices.}
    \label{weightq2}
    \end{figure*}

In the ``Q-SSE" method, there are two same types of updates as that of the ``J-SSE" method. The first update (diagonal update) is the same as the general one, where only the weight are changed. The weights of the $J$ vertices don't contain the magnetic field and instead the weights of the $Q$ vertices contain the magnetic field. The weights for the $Q$ operators, which have $36$ kinds of vertices, are shown in Fig. (\ref{weightq2}). In this figure, we just show the absolute value of weight and ignore the minus sign that may appear in the matrix elements of off-diagonal operators. 
This is because for the $JQ_2$ model in bipartite lattice, the number of the off-diagonal operators with negative weight is required to be even in every allowed configuration in the SSE method, in order to satisfy the ``imaginary time" periodicity (it is just the requirement of the trace of the partition function)\cite{sse-loop}. 

 The second update is also the directed-loop update (off-diagonal update) but it will be different from the ``J-SSE" method. When a loop encounters the $J$ operators, it will just do the switch-and-reverse move. However, when the loop encounters the $Q$ operators, we need solve the directed-loop equations for the $Q$ operators and using these solutions to do the directed-loop update. 

The updates for the ``JQ-SSE" method are similar. As both the $J$ and $Q$ operators contain the magnetic field, the weights for the diagonal $J$ and $Q$ operators should contain part of magnetic field that are assigned to these operators. The first update (diagonal update) is the same as previous and only the weights are different. However for the second update (directed-loop update), both $J$ and $Q$ operators should do the directed-loop update based on the solutions of the directed-loop equations for $J$ and $Q$ cases. The solutions for the $J$ operators can be found in Eq. ($29$),($31$) in Ref. \cite{sse-direct} and the solutions for the $Q$ operators will be shown later in this section.


In Fig. (\ref{confq2}) we have shown all the types of the $Q$ vertices. Every $Q$ vertex can be thought of containing left vertex and right vertex. Both vertices (left vertex and right vertex) are similar to the $J$ vertices. Because of symmetry reasons (spins/bonds permuting and imaginary time inversion), we only need to consider the loop appearing on the left vertex. If the loop appears on the left vertex,  there are two independent sets of directed-loop equations for every type of right vertex. These two sets of equations are similar to the equations appearing in the ``J-SSE" method. As the right vertex has $6$ different types, the total number of independent sets of directed-loop equations is $2\cdot6=12$. We have not excluded the cases which vertices weight are zero. The weights of these vertices may not be zero if we choose another constant rather than $2h_q$.

  \begin{figure}[t]
  \includegraphics[width=75mm,clip]{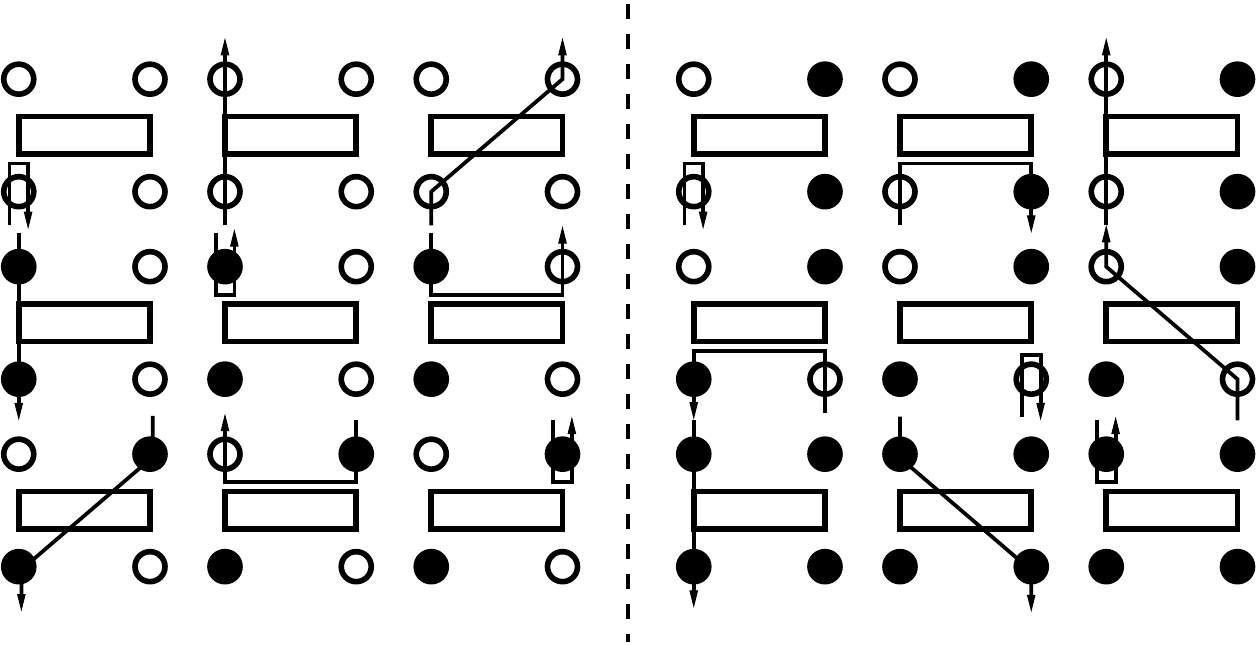}
  \caption{Two independent assignments of directed-loop segments for every column in Fig. (\ref{confq2}). All other directed-loop segments can be derived from these two segments by certain symmetry transformation. The lines with arrows are the directed loops and the arrows present the direction of loops.}
  \label{twoset}
  \end{figure}

Now we try to solve these $12$ sets of equations. For every column in Fig. (\ref{weightq2}), there are two independent assignments of directed-loop segments, as shown in Fig. (\ref{twoset}). These two assignments correspond two independent sets of directed-loop equations. All other assignments can be derived from these two assignments by certain symmetry transformation. As every set of equations should obeys
\be
\sum_xW(s,e,x)=W_s\label{eq:set}
\ee
where $s$ denotes the configuration of a vertex, which has a weight $W_s$. $W(s,e,x)\equiv W_sP(s,e\rightarrow s^\prime,x)$, $e$ is the entrance leg of the vertex and $x$ is the exit leg of the vertex. $P(s,e\rightarrow s^\prime,x)$ means that if the entrance leg is $e$ in a vertex with a configuration $s$, the loop will exit the vertex from leg $x$ with a probability $P(s,e\rightarrow s^\prime,x)$. $s^\prime$ is the new configuration of this vertex after the loop going through this vertex. According to Eqs. (\ref{eq:set}), we can get the corresponding directed-loop equations for every set and the solutions can be derived easily. Note that every set of directed-loop equations has an infinite number of solutions, the solution shown below is particular: we minimize the bounce probability (the entrance leg and exit leg are the same). Such idea is based on the intuitive hypothesis (we have no rigorous proof) that minimizing the bounce probability will increase the simulation efficiency. It can not be ruled out that there exists a more efficient directed-loop solution in which the bounce probability is not minimized.

Now we present the solutions of all the $12$ sets of equations with minimized bounce probability. We will show how to solve the directed-loop equations step-by-step in Appendix. \ref{app:solu}. Firstly, we present the solutions of the two sets of equations for the first column in Fig. (\ref{weightq2}), where the right vertex is $\Gamma_1$. The left assignments of directed-loop segments, shown in Fig. (\ref{twoset}), give the first set of equations: 
\be
4h_q&=&b_1 + a + b\nonumber\\
3h_q&=&a + b_2 + c\label{eq:set1}\\
0&=&b + c + b_3\nonumber
\ee
where the left-hand sides are the vertex weights in the spin configuration space and those on the right are weights in the enlarged configuration space of spins and directed-loop segments. The probabilities of selecting the exit leg are dividing the weights in the extended configuration space by the weight of the bare vertex (spin configuration space). For example, if the loop encounters a vertex $\Gamma_1$ and the entrance leg is the lower left leg (the first row of left part in Fig. (\ref{twoset})), the probability of choosing the lower left leg as exit leg is $b_1/4h_q$, the probability of choosing the lower right leg as exit leg is $0$, the probability of choosing the upper left leg as exit leg is $a/4h_q$ and the probability of choosing the upper right leg as exit leg is $b/4h_q$. The summation of the four probabilities is $1$. It is the same for other directed-loop segments and only the bare weights and weights in the extended space are different.

The solution of this set of equations is
\be
b_1&=&h_q\qquad b_2=0\qquad b_3=0\nonumber\\
a&=&3h_q \qquad b=0  \qquad c=0
\label{solu:1}
\ee

The right assignments give the second set of equations:
\be
3h_q=b_1^\prime+a^\prime+b^\prime\nonumber\\
0=a^\prime+b_2^\prime+c^\prime\label{eq:set2}\\
2h_q=b^\prime+c^\prime+b_3^\prime\nonumber
\ee
The solution can be
\be
b_1^\prime&=&h_q\qquad b_2^\prime=0\qquad b_3^\prime=0\nonumber\\
 a^\prime&=&0 \qquad b^\prime=2h_q  \qquad c^\prime=0
\label{solu:2}
\ee

Secondly, the two sets of equations for the second and third columns in Fig. (\ref{weightq2}), where the right vertices are $\Gamma_2$ and $\Gamma_3$ respectively, are the same. The left assignments give a set of equations:
 \be
 3h_q&=&b_1 + a + b\nonumber\\
 2h_q+\frac{1}{4}&=&a + b_2 + c\label{eq:set3}\\
 \frac{1}{4}&=&b + c + b_3\nonumber
 \ee
The solution is 
 \be
 {\rm if }(h_q\leq \frac{1}{2})\nonumber\\
 b_1&=&0\qquad b_2=0\qquad b_3=0\nonumber\\
 a&=&\frac{5}{2}h_q \qquad b=\frac{h_q}{2}  \qquad c=\frac{1}{4}-\frac{h_q}{2}\nonumber\\
 \nonumber\\
 {\rm if }(h_q>\frac{1}{2})\nonumber\\
 b_1&=&h_q-\frac{1}{2}\qquad b_2=0\qquad b_3=0\nonumber\\
  a&=&2h_q+\frac{1}{4} \qquad b=\frac{1}{4}  \qquad c=0  
\label{solu:3}
 \ee

The equations of the right assignment for these two columns are
 \be
 2h_q+\frac{1}{4}=b_1^\prime+a^\prime+b^\prime\nonumber\\
 \frac{1}{4}=a^\prime+b_2^\prime+c^\prime\label{eq:set4}\\
 h_q=b^\prime+c^\prime+b_3^\prime\nonumber
 \ee
 The solution can be
 \be
 b_1^\prime&=&h_q\qquad b_2^\prime=0\qquad b_3^\prime=0\nonumber\\
  a^\prime&=&\frac{1}{4} \qquad b^\prime=h_q  \qquad c^\prime=0
\label{solu:4}
 \ee

Thirdly, the two sets of equations for the fourth and fifth columns in Fig. (\ref{weightq2}) are also the same, where the right vertices are $\Gamma_4$ and $\Gamma_5$. The equations for the left set are
 \be
 0&=&b_1 + a + b\nonumber\\
 \frac{1}{4}&=&a + b_2 + c\label{eq:set5}\\
\frac{1}{4} &=&b + c + b_3\nonumber
 \ee
 The solution can be
 \be
 b_1&=&0\qquad b_2=0\qquad b_3=0\nonumber\\
 a&=&0 \qquad b=0  \qquad c=\frac{1}{4}
\label{solu:5}
 \ee
 The equations of the right set are
 \be
 \frac{1}{4}=b_1^\prime+a^\prime+b^\prime\nonumber\\
 \frac{1}{4}=a^\prime+b_2^\prime+c^\prime\label{eq:set6}\\
 0=b^\prime+c^\prime+b_3^\prime\nonumber
 \ee
 The solution can be
 \be
 b_1^\prime&=&0\qquad b_2^\prime=0\qquad b_3^\prime=0\nonumber\\
  a^\prime&=&\frac{1}{4} \qquad b^\prime=0  \qquad c^\prime=0
\label{solu:6}
 \ee

At last, we give the two sets of equations for the sixth column in Fig. (\ref{weightq2}) and the right vertex is $\Gamma_6$. The equations for the left set are
 \be
 2h_q&=&b_1 + a + b\nonumber\\
 h_q&=&a + b_2 + c\label{eq:set7}\\
0 &=&b + c + b_3\nonumber
 \ee
 The solution can be
 \be
 b_1&=&h_q\qquad b_2=0\qquad b_3=0\nonumber\\
 a&=&h_q \qquad b=0  \qquad c=0
\label{solu:7}
 \ee
 The right set equations are:
 \be
 h_q=b_1^\prime+a^\prime+b^\prime\nonumber\\
 0=a^\prime+b_2^\prime+c^\prime\label{eq:set8}\\
 0=b^\prime+c^\prime+b_3^\prime\nonumber
 \ee
 The solution can be
 \be
 b_1^\prime&=&h_q\qquad b_2^\prime=0\qquad b_3^\prime=0\nonumber\\
  a^\prime&=&0 \qquad b^\prime=0  \qquad c^\prime=0
\label{solu:8}
 \ee

Based on these solutions, we can construct the directed-loop update. If we only use these solutions shown above, where the magnetic field is combined to the $Q$ operators, we can get the ``Q-SSE" method. However if we not only use the above solutions but also use the solutions that the magnetic field are combined to the $J$ operators, we can get the ``JQ-SSE". In this case, we need divide the magnetic field into two parts. One is put into the $J$ operators and the second is put into the $Q$ operators. In the next section, we will present the simulations results of the ``Q-SSE" method and the ``JQ-SSE" method.

   \begin{figure}[t]
   \includegraphics[width=75mm,clip]{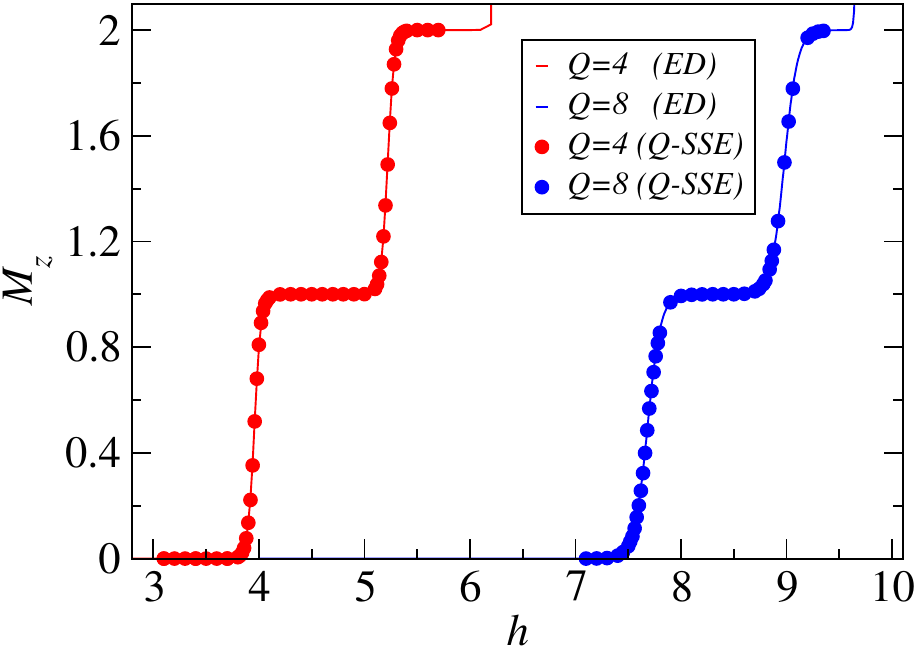}
\caption{Magnetization properties versus magnetic field for the $JQ_2$ model on square lattice. The lattice size is $4\times4$ and the inverse temperature is $\beta=32$. The solid circles are results of the ``Q-SSE" method and the solid lines are results of the ``ED" method.}
    \label{fig:ED-L4}
    \end{figure}
   \begin{figure}[t]
   \includegraphics[width=75mm,clip]{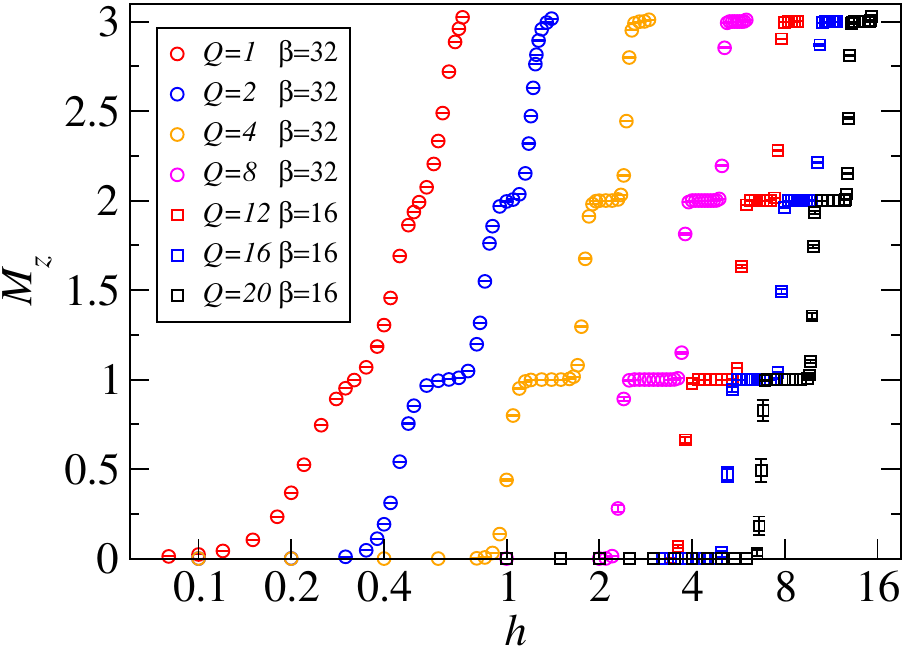}
  \caption{Simulation results of the ``Q-SSE" method on $16\times16$ square lattice at $\beta=16$ or $32$.}
   \label{fig:qsse-resu}
     \end{figure}

\section{Simulations Results}
\label{sec:simu}
In this section, we present the simulation results of the ``Q-SSE" and ``JQ-SSE" methods. We will firstly show that the modified ``Q-SSE" method is correct and the proof of correctness of the ``JQ-SSE" method is shown in Appendix.\ref{app:jqsse}. Then we compare the efficiency of the three different SSE methods.

\subsection{Correctness of the ``Q-SSE" method}
In this subsection, we will prove the correctness of the ``Q-SSE" method introduced above. As our modified program focuses on the properties of models with magnetic field, we mainly concentrate on the magnetization, which are defined as
\be
M_z=\sum_{i=1}^NS_i^z.
\ee

In Fig. (\ref{fig:ED-L4}), we show the simulation results of the $JQ_2$ model on square lattice with $Q=4$ and $8$. The system size is $4\times4$ with inverse temperature $\beta=32$. In this figure, the magnetization properties with magnetic field $h$ for both the ``Q-SSE" method and the ED method coincide. The results of these two different methods do agree with each other and it certificates the correctness of the ``Q-SSE" method.

After proving the correctness of the ``Q-SSE" method, we think the ``JQ-SSE" method will also be right, which is just the combination of the ``J-SSE" and ``Q-SSE" method. In Appendix. \ref{app:jqsse}, we present the results of the ``JQ-SSE" method, which prove the correctness of the ``JQ-SSE" method.

\subsection{Efficiency of the ``Q-SSE" method}
 \begin{figure}[t]
 \includegraphics[width=41.1mm,clip]{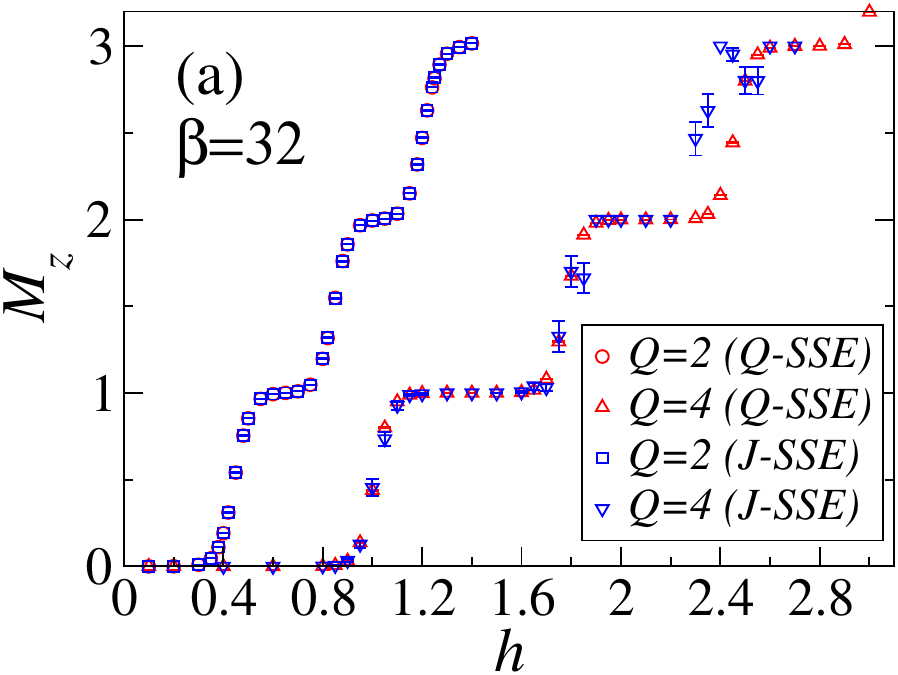}
 \includegraphics[width=43.2mm,clip]{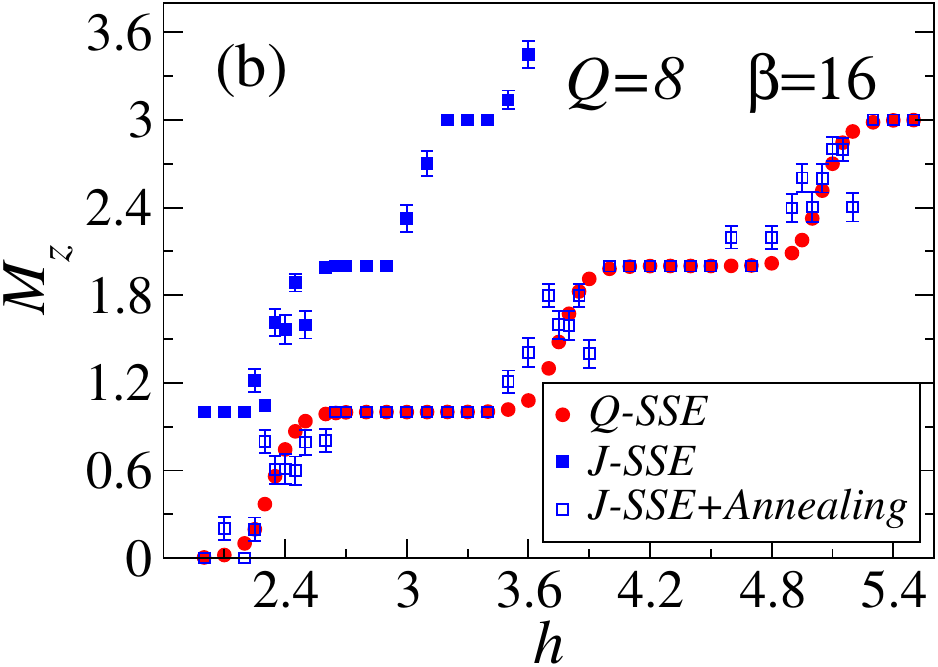}
 \includegraphics[width=41.5mm,clip]{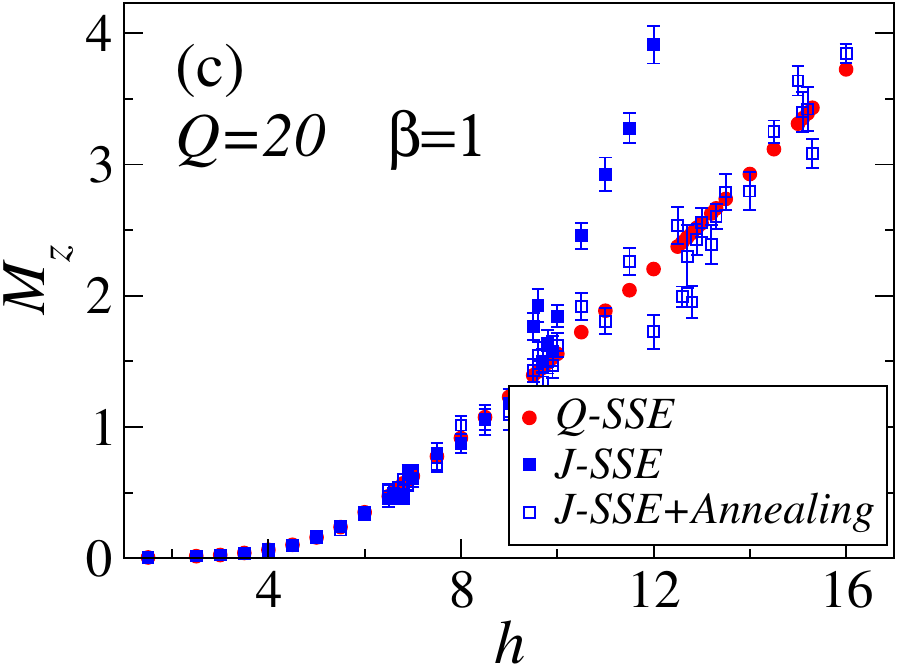}
 \includegraphics[width=41.5mm,clip]{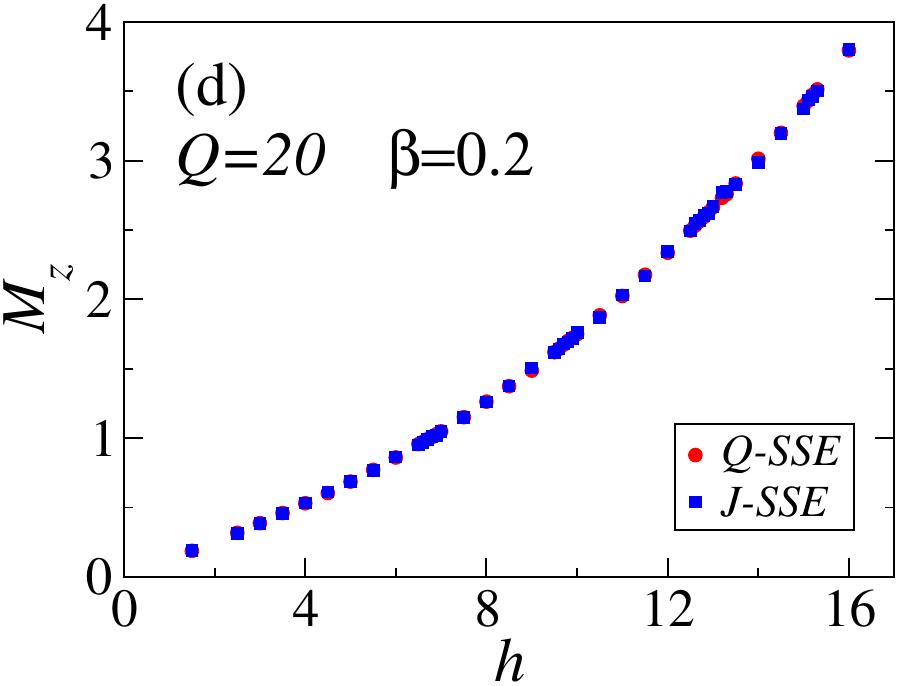}
 \caption{Simulation results for magnetization versus external magnetic field with (a) $Q=2,4$ with $\beta=32$; (b) $Q=8$ with $\beta=16$; (c) $Q=20$ with $\beta=1$; and (d) $Q=20$ with $\beta=0.2$.}
 \label{fig:compare-SSE}
 \end{figure}

In this subsection, we focus on the efficiency of ``Q-SSE" method.
We firstly present the results of the ``Q-SSE" method on $16\times16$ square lattice with $Q=1,2,4,8,12,16,20$ at $\beta=32$ or $16$ in Fig. \ref{fig:qsse-resu}. The quality of magnetization curves is really good. The step structure of magnetization is really clear and the error bars are almost smaller than the symbol size. From this figure, we can see that we need larger external magnetic field to change the magnetization for larger $Q$ interactions.

As mentioned earlier in this paper, the ``J-SSE" method will become more and more worse when $Q$ becomes larger as the autocorrelation time increases significantly. Fig. (\ref{fig:compare-SSE}) shows the simulation results of both the ``J-SSE" and ``Q-SSE" methods. We use the same Monte Carlo parameters for both methods ($50000$ MCSs for equilibration, $50000\times20$ MCSs for measurement, we fix the number of loops in the directed-loop update in one MCS). As seen in Fig. (\ref{fig:compare-SSE})(a), when $Q$ is small ($Q=2$), both the ``J-SSE" method and the ``Q-SSE" method give the correct results for magnetization for a low temperature $\beta=32$. However, when $Q=4$, we can find the results of the ``J-SSE" method become worse, the fluctuations of data become larger, especially when $M_z>2$. The quality of results for the ``Q-SSE" method is still good enough. The results for $Q=8$ at $\beta=16$ are shown in Fig. (\ref{fig:compare-SSE})(b). For this large value of $Q$, even the $\beta$ become small, the ``J-SSE" method gives incorrect results in our simulations. It is because the autocorrelation time is really large for the ``J-SSE" method, the independent configurations change very slowly. For finite MCSs, the configurations may not be thermalized or we can not get enough independent configurations if we start with a thermalized configuration. Of course we can use other optimization methods in the ``J-SSE" method to improve the results, such as the replica exchange method and the annealing method \cite{para,annel}. In this paper, we adopt the annealing method in the ``J-SSE" method, which slowly reduces the temperature. We denote this method as ``J-SSE+Annealing". After annealing, the magnetizations give the right results. But the error bars are still large. However, the quality of the results in ``Q-SSE" method is as good as that of small $Q$. In Fig. (\ref{fig:compare-SSE})(c), we present the results for $Q=20$ at $\beta=1$. The value of $Q$ is really large and the temperature is really high. However, even for this high temperature, the results of the ``J-SSE" method only keep correct for small magnetic field $h$. While the results for large field are still incorrect. After applied the annealing technique, the results will be right but the error bars are also very large. If we raise the model to a much more higher temperature $\beta=0.2$, the results of the ``J-SSE" method are finally correct for all calculated magnetic fields, as shown in Fig. (\ref{fig:compare-SSE})(d).

\begin{figure}[t]
\includegraphics[width=80mm,clip]{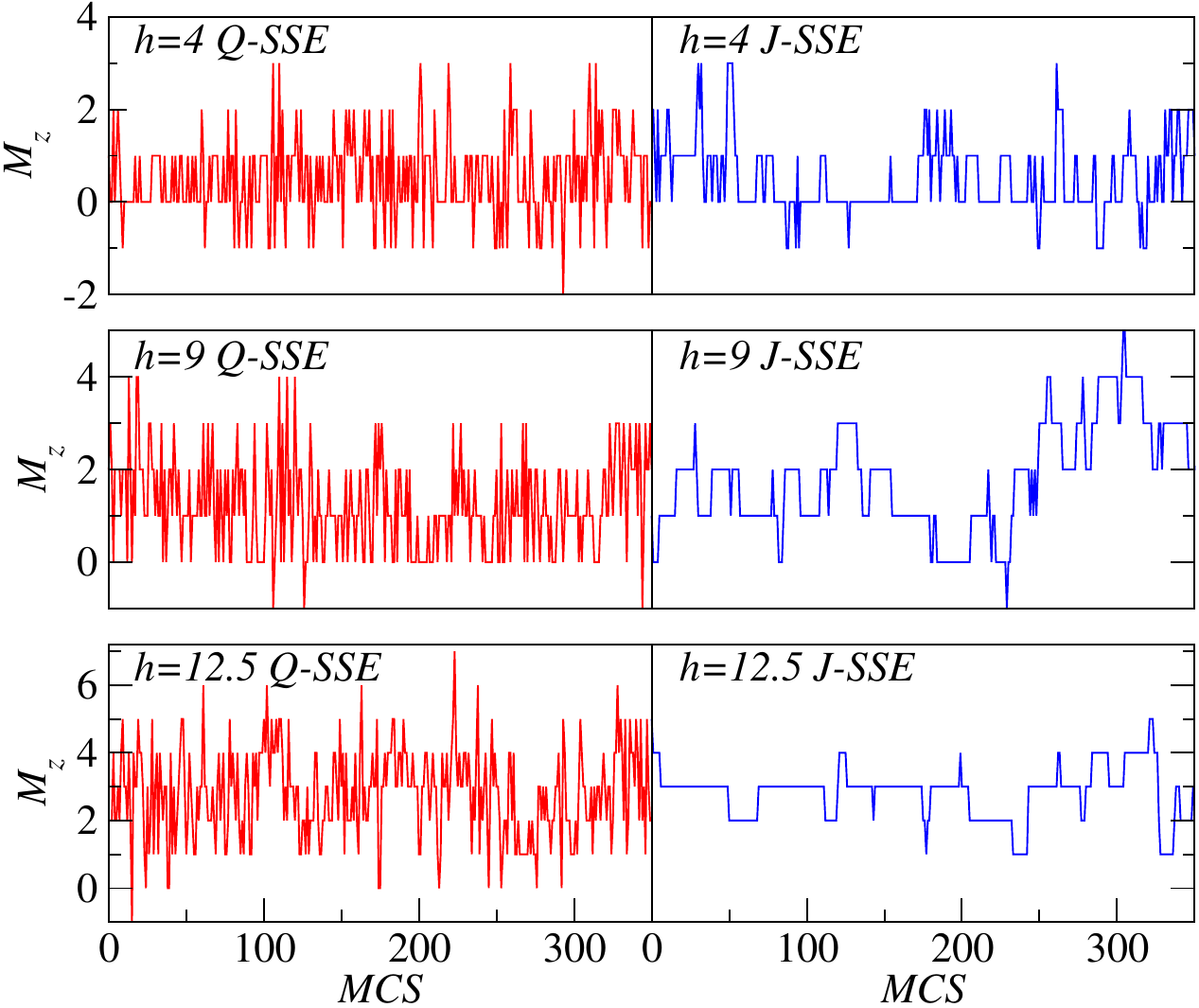}
\caption{Typical evolutions of magnetization for the $JQ_2$ model with $Q=20, \beta=0.2$. The magnetic fields are $h=4, 9, 12.5$.}
    \label{evolu}
    \end{figure}

From Fig. (\ref{fig:compare-SSE}), we can clearly see that the general ``J-SSE" method will fail for large $Q$ and low temperature with finite MCSs. In order to elucidate this conclusion in more detail, we present the typical evolutions for both the ``J-SSE" method and the ``Q-SSE" method in Fig. (\ref{evolu}). We also present the autocorrelation properties later. Fig. (\ref{evolu}) shows the typical evolutions of magnetization with $350$ MCSs for both the ``J-SSE" method and the ``Q-SSE" method at $Q=20$ on $16\times16$ square lattice. The inverse temperature chosen in this figure is $\beta=0.2$. We choose three different magnetic fields $h=4,9,12.5$ and the expectation values of the magnetization for these fields are around $0.5, 1.5, 2.5$ respectively (it can be seen in Fig. (\ref{fig:compare-SSE})(d)). We can clearly see that the evolutions of magnetization is faster in the ``Q-SSE" method than that in the ``J-SSE" method. When the external field increases, the evolution in the ``J-SSE" method will become even slower and it seems the evolution speed in the ``Q-SSE" method changes very little. In this part, we have not shown the efficiency of the ``JQ-SSE" method. It will be presented in the next section via the autocorrelation times.  

     \begin{figure}[t]
    \includegraphics[width=75mm,clip]{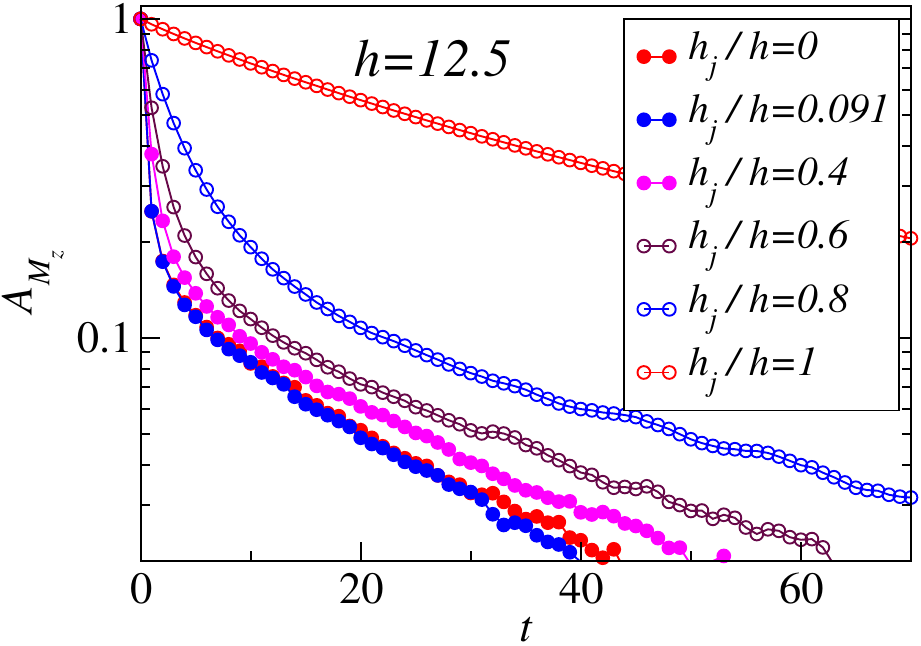}
\caption{The normalized autocorrelation function for $Q=20$, $h=12.5$ and $\beta=0.2$ on $16\times16$ square lattice.}
     \label{fig:cor}
       \end{figure}

\subsection{Autocorrelations}
  \begin{figure}[t]
  \includegraphics[width=40.2mm,clip]{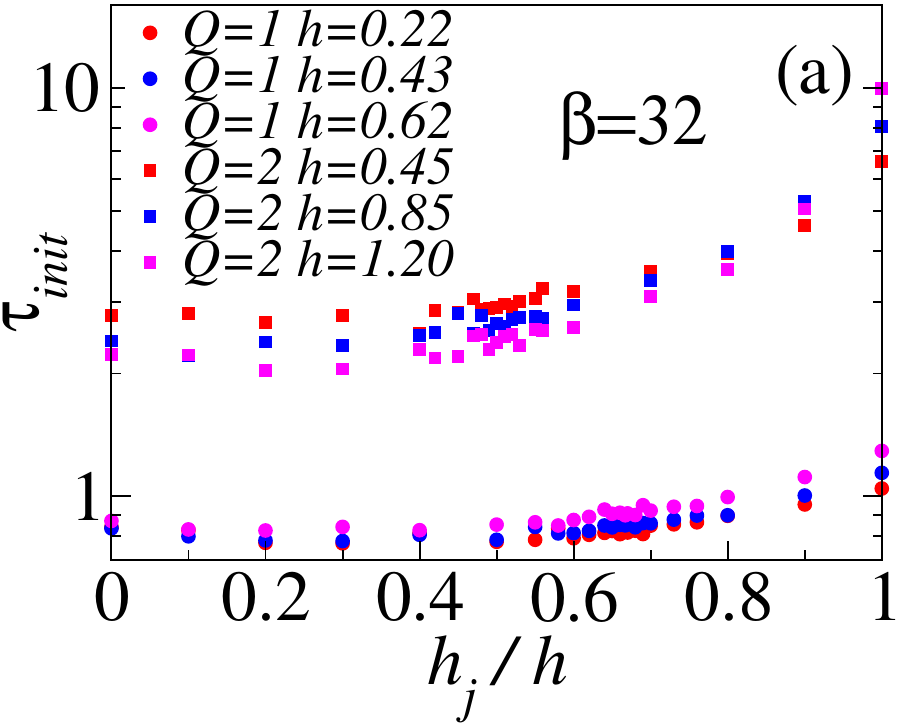}
  \includegraphics[width=42.7mm,clip]{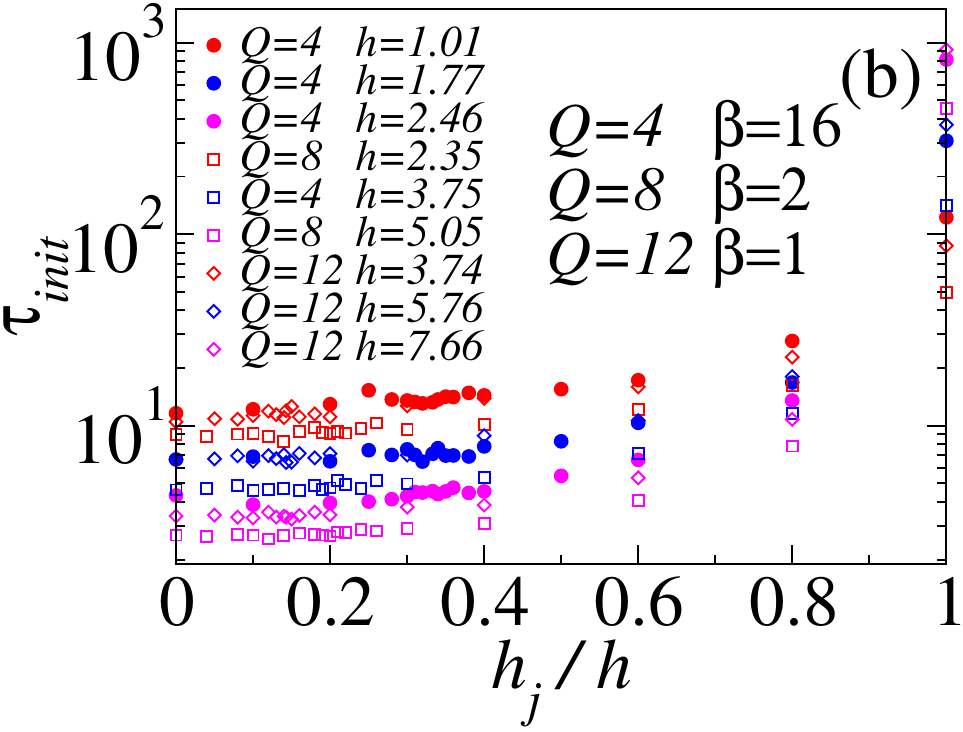}
  \includegraphics[width=41.5mm,clip]{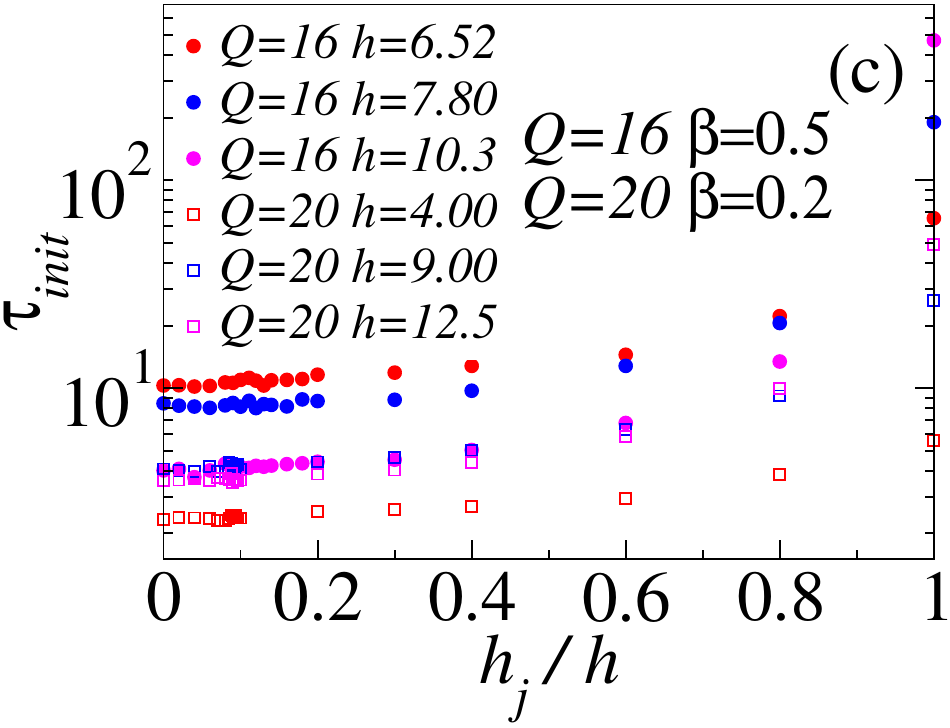}
  \includegraphics[width=42.5mm,clip]{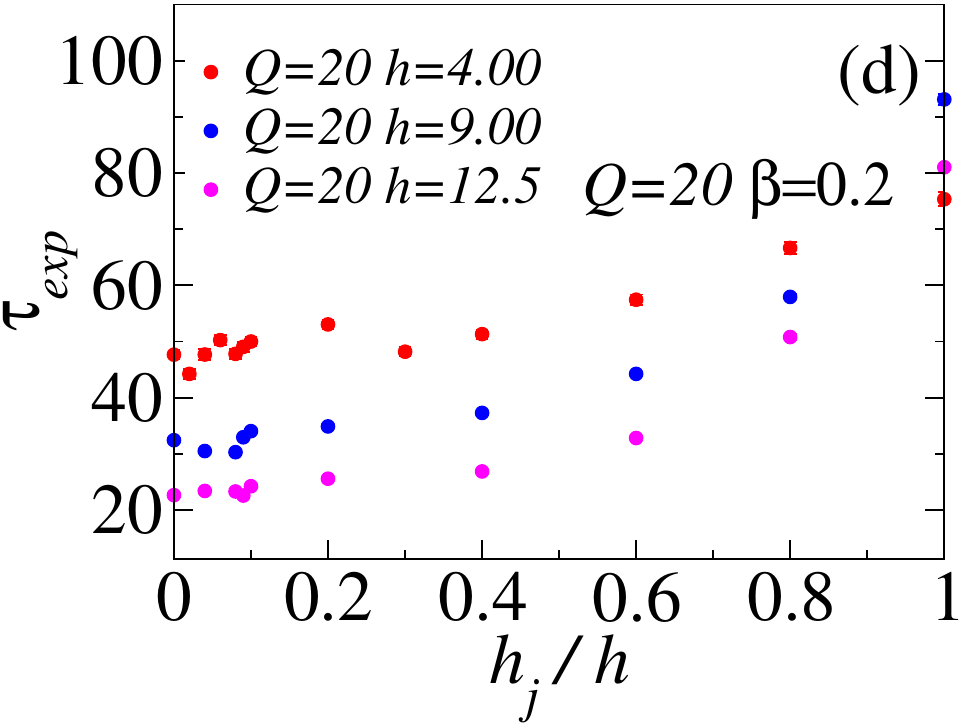}
 \caption{Integrated autocorrelation time and exponential autocorrelation time of magnetizations versus the division ratio of magnetic field in the ``JQ-SSE" method for $JQ_2$ model on $16\times6$ square lattice. (a-c) show the integrated autocorrelation for $Q=1,2$, $Q=4,8,12$ and $Q=16,20$. (d) shows the exponential autocorrelation time for $Q=20$.}
  \label{fig:time}
  \end{figure}

The autocorrelation functions provide direct quantitative measurement of the efficiency of a Monte Carlo method in generating the independent configurations. In this part, we will focus on the autocorrelations. For a quantity $O$, the normalized autocorrelation function is defined as
\be
A_O(t)=\frac{\la O(i+t)O(i)\ra-\la O(i)\ra^2}{\la O(i)^2\ra-\la O(i)\ra^2}
\ee
where $i$ and $t$ are Monte Carlo times (we use the unit of $1$ MCS). The brackets indicate the average over time $i$. For large time separations, the autocorrelation function decays exponentially as
\be
A(t)\xrightarrow{t\rightarrow\infty}a{\rm e}^{-t/\tau_{exp}}
\label{eq:exp}
\ee 
$\tau_{exp}$ is the exponential autocorrelation time and $a$ is a constant. This time is given by the slowest mode of the simulation to which the observable $O$ couples. At smaller time, usually other modes contribute and $O(t)$ behaves no longer purely exponentially.

 Here we also introduce another time: the integrated autocorrelation time, which is defined as
\be
\tau_{int}[O]=1/2+\sum_{t=1}^{\infty}A_O(t).
\label{eq:int}
\ee
This time is the autocorrelation measure of the greatest practical utility \cite{int-cor}. In general, these two times are different. Only if $A(t)$ is a pure exponential, the two times coincide. In this paper, we focus on the autocorrelations of magnetizations. 

In this part, we not only pay attention to the autocorrelations of magnetizations in the ``J-SSE" and ``Q-SEE" methods, we also study the autocorrelations in the ``JQ-SSE" method. In the ``JQ-SSE" method, as mentioned above, the external magnetic field should be divided into two parts and be combined with $J$ interactions (Heisenberg interactions) and $Q$ interactions respectively. The directed-loop updates will be carried out on both the $J$ and $Q$ bonds. In this method, there is another freedom: the division ratio. For a magnetic field $h$ on a spin, we can divide it into two magnetic fields. We denote the strength of the first field as $h_j$ and this field will be combined to the $J$ interactions. The strength of the second field, which will be put into the $Q$ interactions, is $h-h_j$. The value of $h_j$ should be $0\le h_j\le h$. We define the division ratio of the magnetic field as $h_j/h$, which is the ratio of magnetic field that will be applied to the $J$ interactions. The ``J-SSE" and ``Q-SSE" methods are two extreme cases: $h_j/h=1$ for the ``J-SSE" method, $h_j/h=0$ for the ``Q-SSE" method.

We firstly present the normalized autocorrelation function at $Q=20$, $h=12.5$, $\beta=0.2$ on $16\times16$ square lattice in Fig. (\ref{fig:cor}). We can clearly find when $h_j/h=1$ (``J-SSE" method), the autocorrelation function is the largest. The ratio of the smallest autocorrelation function is close to $0$ (we can not determine the exact ratio).

Next in Fig. (\ref{fig:time}), we present the magnetization integrated autocorrelation time for $Q=1,2,4,8,12,16,20$ and the exponential autocorrelation time for $Q=20$. The integrated autocorrelation times are calculated from Eq. (\ref{eq:int}) and we fit the exponential autocorrelation time from autocorrelation functions based on Eq. (\ref{eq:exp}). In this figure, we choose different temperatures for different $Q$. That is because we need ensure the ``JQ-SSE" methods in all division ratios give the right results in our finite MCSs (we have verified the results for all ratios are correct). In addition, we choose $3$ different magnetic fields for every $Q$, in which the expectation values of magnetization are around $0.5,1,1.5$ respectively. From Fig. (\ref{fig:time}), we can find that the autocorrelation times are the biggest for $h_j/h=1$ (``J-SSE" method) even for $Q=1$ and $2$. When $Q$ increases, the autocorrelation times increase significantly at $h_j/h=1$. It means the ``J-SSE" method is the worst choice to do the simulations of the $JQ_2$ model with magnetic field for these values of $Q$ (Of course, we can argue that when $Q$ is much smaller, the ``J-SSE" method is the best choice). Though the autocorrelation for ``J-SSE" method is the largest for small $Q$, the autocorrelation times for the ``J-SSE" method are still not large ($\tau_{init}<10$ for $Q=1,2$). So the simulations of the ``J-SSE" method for small $Q$ are still good enough. For large $Q$, the smallest autocorrelation times seem to be close to $h_j/h=0$. In addition, for a large range of the ratio around $0$, the autocorrelation times do not change too much.
We can not clearly find the best ratio for large $Q$ in this figure. So we suggest that the ``Q-SSE" method is good enough for simulating the $JQ_2$ model with large $Q$. We do not need to optimize the ratio in the ``JQ-SSE" method to find the smallest autocorrelation time, which is very close to that of $h_j/h=0$. The results shown in Fig. (\ref{fig:time}) are based on $(5-10)\times10^5$ MCSs for each data point.

\section{summary}
\label{sec:summary}
In this paper, we argue that if we study the $JQ_2$ model with external magnetic field, the general SSE method with directed loops (``J-SSE" method) will be good enough for small $Q$ interactions. However, when the $Q$ interactions become large and the temperature is low, this general SSE method may fail with finite MCSs. Here we introduce the modified SSE methods (the ``Q-SSE" and ``JQ-SSE" methods) to deal with this problem. These modified methods can really decrease the autocorrelation times especially for large $Q$ interactions. Thus it can really speed up the simulations. In addition, we argue that when doing simulations of the $JQ_2$ model with large $Q$, the ``Q-SSE" method is good enough. We do not need to optimize the ratio $h_j/h$ to find the smallest autocorrelation time in the ``JQ-SSE" method, which is very close to that of the ``Q-SSE" method.

The principle behind these modified methods is that for the $JQ_2$ model, the products of operators in the SSE configurations $S_M$ not only contain the $J$ operators but also have the $Q$ operators. The general ``J-SSE" method only make use of the $J$ operators to do the directed-loop updates. When $Q$ becomes large, the portion of the $J$ operators in the operator products will be very small. It means the general method only affect really small part of the products. The new methods, introduced here, consider the main part of the operators (``Q-SSE" method) or even all the part of products (``JQ-SSE" method), which are much better than the general ``J-SSE" method.  

Such methods and idea can be applied to other models, such the $JQ_3$ model and the $CBJQ$ model. If there are $N$ types of interactions, we can also divide the magnetic field into $N$ parts and combine them with every type of interactions respectively. These methods will certainly speed up the simulations of these models.

In this article, we have not paid attention to the effect of the constant added to the Hamiltonian and the different solutions to the directed loop-equations. The three SSE methods all choose the smallest constant and minimize the bounce probability. We focus on the efficiency difference when combining the magnetic fields with different types of operators. The two new treatments of the magnetic field improve the efficiency significantly.

\begin{acknowledgments}
This work is supported by Beijing Institute of Technology Research Fund Program for Young Scholars and the National Natural Science Foundation of China under Grants No. 12304171.
\end{acknowledgments}

\appendix

\section{Solutions of the directed-loop equations}
\label{app:solu}
In this section, we will present how to solve the directed-loop equations in Eqs. (\ref{eq:set1})(\ref{eq:set2})(\ref{eq:set3})(\ref{eq:set4})(\ref{eq:set5})(\ref{eq:set6})(\ref{eq:set7})(\ref{eq:set8}) step-by-step. The solution for the general form of directed-loop equations has been discussed in Ref. \cite{sse-direct}. In the $JQ_2$ model, there are $3$ equations in every set. It is convenient to label three weights as $W_1$, $W_2$, $W_3$ and $W_3\geq W_2\geq W_1$. The weights in every set can be relabeled in this order and every set of equations can be written as
\be
W_1&=&=a_{11}+a_{12}+a_{13}\nonumber\\
W_2&=&=a_{21}+a_{22}+a_{23}\nonumber\\
W_3&=&=a_{31}+a_{32}+a_{33}
\ee
where all $a_{ij}$ should be non-negative and $a_{ij}=a_{ji}$. The bounce probabilities are determined by $a_{ii}, i=1,2,3$. As we want to minimize the bounce probabilities, we can set $a_{ii}=0$ for $i=1,2,3$. Then there are three independent unknowns ($a_{12}, a_{13}, a_{23}$) and three equations, which make the solution unique. However every $a_{ij}$ should be non-negative and we can find only when $W_3\leq W_1+W_2$, all the three $a_{ii}$ can be $0$.
In this condition, the solution is
\be
a_{12}&=&a_{21}=(W_1+W_2-W_3)/2\nonumber\\
a_{13}&=&a_{31}=(W_1-W_2+W_3)/2\nonumber\\
a_{23}&=&a_{32}=(-W_1+W_2+W_3)/2
\label{solu:A1}
\ee  

When $W_3>W_1+W_2$, we can always permit one bounce probabilities not be zero (this is the bounce in the largest weight $W_3$ configuration). We set $a_{11}=a_{22}=0$ and $a_{33}=W_3-W_1-W_2$ and the solution for the three unknowns is
\be
 a_{12}&=&a_{21}=0\nonumber\\
 a_{13}&=&a_{31}=W_1\nonumber\\
 a_{23}&=&a_{32}=W_2
\label{solu:A2}
\ee
Based on Eqs. (\ref{solu:A1})(\ref{solu:A2}), we now solve the $8$ independent sets of equations for the $JQ_2$ model with magnetic field.

The weights of the first set of equations (Eqs. (\ref{eq:set1})) are $W_1=0, W_2=3h_q, W_3=4h_q$ (we have relabeled the order of the weight). As $W_1+W_2<W_3$, we can get the solution according to Eqs. (\ref{solu:A2}) 
\be
a_{11}&=&a_{22}=0\nonumber\\
a_{33}&=&W_3-W_1-W_2=h_q\nonumber\\
a_{12}&=&=a_{21}=0\nonumber\\
a_{13}&=&a_{31}=W_1=0\nonumber\\
a_{23}&=&=a_{32}=W_2=3h_q.
\ee
This is just the solution shown in Eqs. (\ref{solu:1}).

The weights of the second set of equations (Eqs. (\ref{eq:set2})) are $W_1=0, W_2=2h_q, W_3=3h_q$. As $W_1+W_2<W_3$, we get the solution according to Eqs. (\ref{solu:A2})
\be
a_{11}&=&a_{22}=0\nonumber\\
a_{33}&=&W_3-W_1-W_2=h_q\nonumber\\
a_{12}&=&a_{21}=0\nonumber\\
a_{13}&=&a_{31}=W_1=0\nonumber\\
a_{23}&=&a_{32}=W_2=2h_q.
\ee
This is just the solution shown in Eqs. (\ref{solu:2}).

The weights of the third set of equations (Eqs. (\ref{eq:set3})) are $3h_q, 2h_q+1/4, 1/4$. The order of the weight depends on the value of $h_q$. When $h_q\leq 1/12$, $W_1=3h_q, W_2=1/4, W_3=2h_q+1/4$. As $W_1+W_2\geq W_3$, we get the solution according to Eqs. (\ref{solu:A1})
\be
a_{11}&=&a_{22}=a_{33}=0\\
a_{12}&=&a_{21}=(W_1+W_2-W_3)/2=h_q/2\nonumber\\
a_{13}&=&a_{31}=(W_1-W_2+W_3)/2=5h_q/2\nonumber\\
a_{23}&=&a_{32}=(-W_1+W_2+W_3)/=1/4-h_q/2\nonumber,\label{eq:A3-1}
\ee
When $1/12\leq h_q\leq 1/4$, $W_1=1/4, W_2=3h_q, W_3=2h_q+1/4$. As $W_1+W_2\geq W_3$, we get the solution according to Eqs. (\ref{solu:A1})
\be
a_{11}&=&a_{22}=a_{33}=0\\
a_{12}&=&a_{21}=(W_1+W_2-W_3)/2=h_q/2\nonumber\\
a_{13}&=&a_{31}=(W_1-W_2+W_3)/2=1/4-h_q/2\nonumber\\
a_{23}&=&a_{32}=(-W_1+W_2+W_3)/=5h_q/2\nonumber,\label{eq:A3-2}
\ee
When $1/4<h_q\leq 1/2$, $W_1=1/4, W_2=2h_q+1/4, W_3=3h_q$. As $W_1+W_2\geq W_3$, we get the solution according to Eqs. (\ref{solu:A1})
\be
a_{11}&=&a_{22}=a_{33}=0\\
a_{12}&=&a_{21}=(W_1+W_2-W_3)/2=1/4-h_q/2\nonumber\\
a_{13}&=&a_{31}=(W_1-W_2+W_3)/2=h_q/2\nonumber\\
a_{23}&=&a_{32}=(-W_1+W_2+W_3)/2=5h_q/2\nonumber\label{eq:A3-3},
\ee
When $h_q>1/2$, $W_1=1/4, W_2=2h_q+1/4, W_3=3h_q$. As $W_1+W_2< W_3$, we get the solution according to Eqs. (\ref{solu:A2})
\be
a_{11}&=&a_{22}=0\\
a_{33}&=&W_3-W_1-W_2=h_q-1/2\nonumber\\
a_{12}&=&a_{21}=0\nonumber\\
a_{13}&=&a_{31}=W_1=1/4\nonumber\\
a_{23}&=&a_{32}=W_2=2h_q+1/4\nonumber
\ee
There are four solutions for the third set of equation under different conditions. But one can find that Eqs. (\ref{eq:A3-1})(\ref{eq:A3-2})(\ref{eq:A3-3}) give the same solution. That is because under these three conditions, the weights all have the property $W_1+W_2\geq W_3$, which means $a_{ii}=0$. Thus the solution of the unknown is unique. We set $W_1=1/4, W_2=2h_q+1/4, W_3=3h_q$ and summarize the solutions for Eqs. (\ref{eq:set3}):
when $h_q\leq1/2$
\be
a_{11}&=&a_{22}=a_{33}=0\nonumber\\
a_{12}&=&a_{21}=1/4-h_q/2\nonumber\\
a_{13}&=&a_{31}=h_q/2\nonumber\\
a_{23}&=&a_{32}=5h_q/2\nonumber
\ee
when $h_q>1/2$
\be
a_{11}&=&a_{22}=0\nonumber\\
a_{33}&=&h_q-1/2\nonumber\\
a_{12}&=&a_{21}=0\nonumber\\
a_{13}&=&a_{31}=1/4\nonumber\\
a_{23}&=&a_{32}=2h_q+1/4
\ee
 This is just the solution shown in Eqs. (\ref{solu:3}).

The weights of the fourth set of equations (Eqs. (\ref{eq:set4})) are $W_1=h_q (1/4), W_2=1/4 (h_q), W_3=2h_q+1/4$ when $h_q\leq1/4$ ($h_q>1/4$). As $W_1+W_2<W_3$, we get the solution according to Eqs. (\ref{solu:A2}). The value of $h_q$ only changes the order of $W_1$ and $W_2$, which does not change the solution. We can present the solutions together. We set $W_1=h_q, W_2=1/4, W_3=2h_q+1/4$, the solution for this set of equation is
\be
a_{11}&=&a_{22}=0\\
a_{33}&=&W_3-W_1-W_2=h_q\nonumber\\
a_{12}&=&a_{21}=0\nonumber\\
a_{13}&=&a_{31}=W_1=h_q\nonumber\\
a_{23}&=&a_{32}=W_2=1/4\nonumber
\ee
This is just the solution shown in Eqs. (\ref{solu:4}).

The weights of the fifth and sixth sets of equations (Eqs. (\ref{eq:set5})(\ref{eq:set6})) are same: $W_1=0, W_2=1/4, W_3=1/4$. As $W_1+W_2\geq W_3$, we get the solution according to Eqs. (\ref{solu:A1})
\be
a_{11}&=&a_{22}=a_{33}=0\\
a_{12}&=&a_{21}=(W_1+W_2-W_3)/2=0\nonumber\\
a_{13}&=&a_{31}=(W_1-W_2+W_3)/2=0\nonumber\\
a_{23}&=&a_{32}=(-W_1+W_2+W_3)/2=1/4\nonumber
\ee
This is just the solution shown in Eqs. (\ref{solu:5})(\ref{solu:6}).

The weights of the seventh set of equations (Eqs. (\ref{eq:set7})) are $W_1=0, W_2=h_q, W_3=2h_q$. As $W_1+W_2<W_3$,
we get the solution according to Eqs. (\ref{solu:A2})
\be
a_{11}&=&a_{22}=0\\
a_{33}&=&W_3-W_1-W_2=h_q\nonumber\\
a_{12}&=&a_{21}=0\nonumber\\
a_{13}&=&a_{31}=W_1=0\nonumber\\
a_{23}&=&a_{32}=W_2=h_q\nonumber
\ee
This is just the solution shown in Eqs. (\ref{solu:7}).

The weights of the eighth set of equations (Eqs. (\ref{eq:set8})) are $W_1=0, W_2=0, W_3=h_q$. As $W_1+W_2<W_3$,
 we get the solution according to Eqs. (\ref{solu:A2})
\be
a_{11}&=&a_{22}=0\\
a_{33}&=&W_3-W_1-W_2=h_q\nonumber\\
a_{12}&=&a_{21}=0\nonumber\\
a_{13}&=&a_{31}=W_1=0\nonumber\\
a_{23}&=&a_{32}=W_2=0\nonumber
\ee
This is just the last solution shown in Eqs. (\ref{solu:8}).
\section{The number of the $J$ and $Q$ bonds in the $JQ_2$ model}
\label{app:number}
\begin{figure}[t]
\includegraphics[width=75mm,clip]{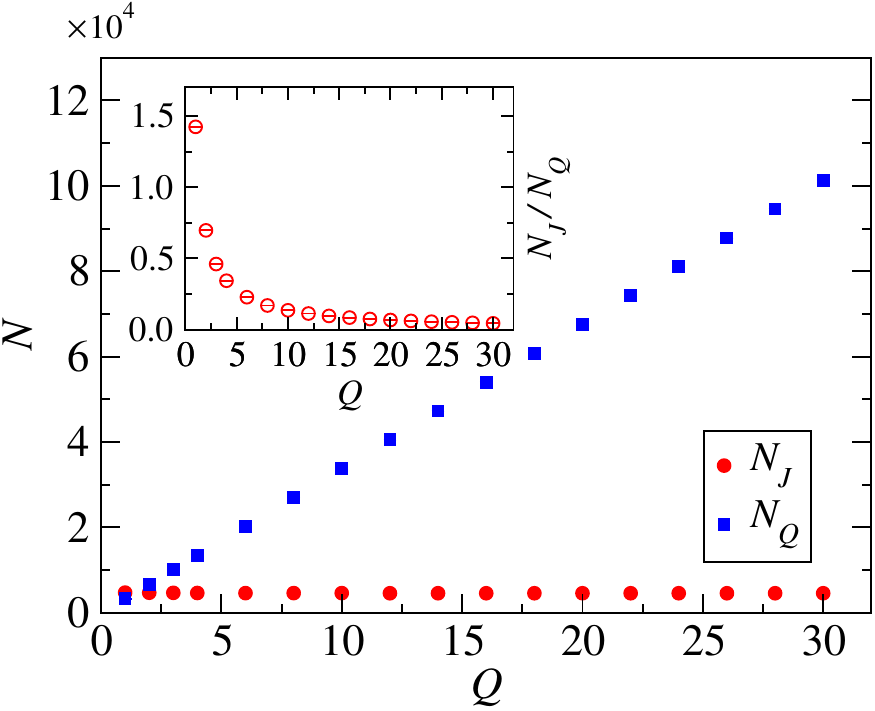}
\caption{The number of the $J$ bonds ($N_J$) and the $Q$ bonds ($N_Q$) in SSE configurations of the $JQ_2$ model without external field. The system size is $L=16$ and the inverse temperature is $\beta=16$. The inset shows the ratio of the number of $J$ bonds to the number of $Q$ bonds.}
\label{fig:bond}
\end{figure}

As mentioned in main text, the $Q$ bonds will appear more frequently than the $J$ bonds in the $JQ_2$ model, when $Q$ is large. In Fig. \ref{fig:bond}, we present simulation results to verify such statement. We perform SSE simulations to study the properties of the number of the $J$ bonds and the $Q$ bonds appearing in the SSE configurations in the $JQ_2$ model without external field. The system is $L=16$ square lattice and the inverse temperature is $\beta=16$. The strength of the $J$ bonds is set to $1$ and the strength of $Q$ bonds ranges from $1$ to $30$. One can find that when the value of $Q$ increases, the number of the $J$ bonds will decrease very slowly and the number of the $Q$ bonds will increase very fast. The inset in Fig. \ref{fig:bond} shows the ratio of the number of $J$ bonds to the number of $Q$ bonds. The ratio will decrease to very small value when $Q$ is very large.

\section{Results for the ``JQ-SSE" method}
\label{app:jqsse}
\begin{figure}[t]
\includegraphics[width=75mm,clip]{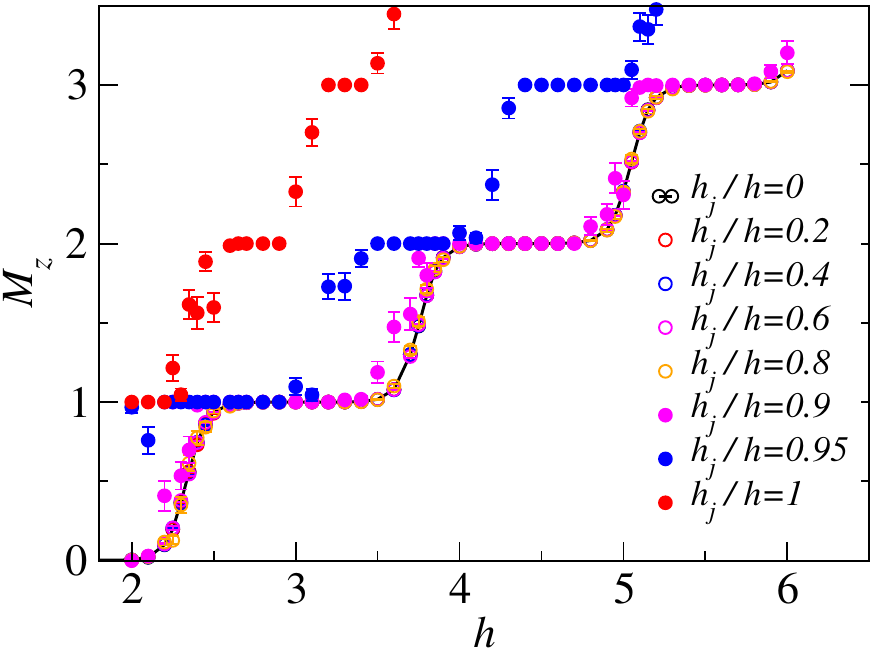}
\caption{Simulation results for magnetization versus external magnetic field with $Q=8$ at $\beta=16$. $h_j/h$ is the ratio of magnetic field applied into $J$ bonds.}
\label{fig:htj}
\end{figure}
In this section, we present the simulation results for the $JQ_2$ model at $Q=8$ with external magnetic field by the ``JQ-SSE" method. The inverse temperature is $\beta=16$ and the system size is $L=16$. $h_j/h$ is the ratio of magnetic field that is applied to the $J$ bonds. We use $50000$ MCSs for equilibration, $50000\times 20$ MCSs for measurement for all $h_j/h$. When $h_j/h=1$, it's just the ``J-SSE" method, which has the largest autocorrelation time. The MCSs used here are not large enough to give the right answer. When $h_j/h=0$, it's the ``Q-SSE" method. We think it gives the right results in our simulations, as we have proven its correctness in main text. In Fig. \ref{fig:htj}, one can find when the ratio $h_j/h\leq 0.8$, the magnetization curves are the same as that of $h_j/h=0$, which means the ``JQ-SSE" method is right. However, when $h_j/h$ is close to $1$, the results diverge from the correct results. The reason is that for ratio close to $1$, the ``JQ-SSE" method is close to the ``J-SSE" method and the most part of magnetic field is still applied into the $J$ bonds. Though the autocorrelation time decreases as the ratio reduces, the autocorrelation times are still large compared to our MCSs. In addition, even for $h_j/h=0.9$, which is close to the ``J-SSE" method, the results are very close to the right results.
\section{Results for the ``Q-SSE" method at large $Q$}

 \begin{figure}[t]
 \includegraphics[width=75mm,clip]{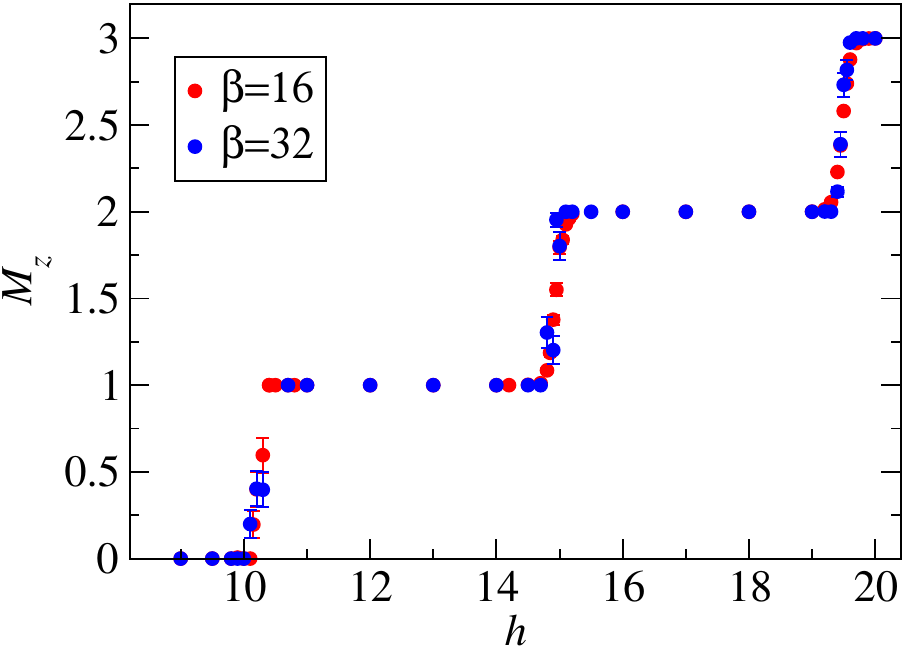}
\caption{Simulation results for magnetization versus external magnetic field with $Q=30$ at $\beta=16,32$}
 \label{fig:q30}
 \end{figure}
In this section, we present simulation results for even large $Q$ strength ($Q=30$) at $\beta=16,32$ via the ``Q-SSE" method. We can clearly see the step structure of magnetization.  
\end{document}